\documentclass[sigconf]{acmart}
\AtBeginDocument{%
  \providecommand\BibTeX{{%
    \normalfont B\kern-0.5em{\scshape i\kern-0.25em b}\kern-0.8em\TeX}}}
    
\copyrightyear{2020} 
\acmYear{2020} 
\setcopyright{acmcopyright}\acmConference[ESEM '20]{ESEM '20: ACM / IEEE International Symposium on Empirical Software Engineering and Measurement (ESEM)}{October 8--9, 2020}{Bari, Italy}
\acmBooktitle{ESEM '20: ACM / IEEE International Symposium on Empirical Software Engineering and Measurement (ESEM) (ESEM '20), October 8--9, 2020, Bari, Italy}
\acmPrice{15.00}
\acmDOI{10.1145/3382494.3410676}
\acmISBN{978-1-4503-7580-1/20/10}

\usepackage[utf8]{inputenc}
\usepackage{diagbox}
\usepackage{color}
\usepackage{multirow}
\usepackage{amssymb}
\usepackage{paralist}
\usepackage{graphicx}
\usepackage{caption}
\usepackage{tcolorbox}
\usepackage{numprint}
\usepackage{url}
\usepackage[inline]{enumitem}
\usepackage{natbib}
\npthousandsep{,}
\usepackage[labelformat=simple]{subcaption}
\newcommand{\rfeat}[1]{\emph{#1}}
\newcommand{\datasetsize}{387}

\newcommand{\etoneabb}{CR}
\newcommand{\ettwoabb}{CD}
\newcommand{\etthreeabb}{S}
\newcommand{\etfourabb}{SR}
\newcommand{\etfiveabb}{SD}
\newcommand{\etsixabb}{SRP}
\newcommand{\etsevenabb}{PGR}
\newcommand{\eteightabb}{PGD}
\newcommand{\etnineabb}{PSR}
\newcommand{\ettenabb}{PSD}
\newcommand{\etelevenabb}{I}

\newcommand{\etone}{Constant Rise}
\newcommand{\ettwo}{Constant Decline}
\newcommand{\etthree}{Stability}
\newcommand{\etfour}{Sudden Rise}
\newcommand{\etfive}{Sudden Decline}
\newcommand{\etsix}{Sudden Rise Plateau}
\newcommand{\etseven}{Plateau Gradual Rise}
\newcommand{\eteight}{Plateau Gradual Decline}
\newcommand{\etnine}{Plateau Sudden Rise}
\newcommand{\etten}{Plateau Sudden Decline}
\newcommand{\eteleven}{Instability}

\title{
On the adoption, usage and evolution of Kotlin features in Android development
}

\author{Bruno Gois Mateus}
\email{Bruno.GoisMateus@etu.uphf.fr}
\affiliation{%
  \institution{Universit\'e Polytechnique Hauts-de-France, France}
}

\author{Matias Martinez}
\email{Matias.Martinez@uphf.fr}
\affiliation{%
  \institution{Universit\'e Polytechnique Hauts-de-France, France}
}

\begin{document}

\begin{abstract}

{\bf Background:} 
Google announced Kotlin as an Android official programming language in 2017, giving developers an option of writing applications using a language that combines object-oriented and functional features.
{\bf Aims:} The goal of this work is to understand the usage of Kotlin features considering four aspects:
\begin{inparaenum}[\it i)]
\item which features are adopted,
\item what is the degree of adoption,
\item when are these features added into Android applications for the first time,
and 
\item how the usage of features evolves along with applications' evolution.
\end{inparaenum}
{\bf Method:} 
Exploring the source code of 387 Android applications, we identify the usage of Kotlin features on each version application's version and compute the moment that each feature is used for the first time.  
Finally, we identify the evolution trend that better describes the usage of these features.
{\bf Results:} 
15 out of 26 features are used on at least 50\% of applications.
Moreover, we found that \rfeat{type inference}, \rfeat{lambda} and \rfeat{safe call} are the most used features.
Also, we observed that the most used Kotlin features are those first included on Android applications.
Finally, we report that the majority of applications tend to add more instances of 24 out of 26 features along with their evolution.
{\bf Conclusions:} Our study 
 generates 7 main findings. We present their implications, which are addressed to developers, researchers and tool builders in order to foster the use of Kotlin features to develop Android applications.

\end{abstract}



\maketitle


\section{Introduction}
\label{sec:introduction}



Currently, the Android platform is the largest mobile platform, with more than 2 million applications published in the official store, Google Play.
Since the first release of Google's mobile operating system, developers have been developing applications mostly using Java and, in some specific scenarios using C++.
However, in 2017, when Google announced Kotlin as an Android official programming language, developers gained another option to write applications.
Moreover, in 2019, Google announced that Android development would become increasingly `Kotlin-first', which means that new APIs and features will be offered first in Kotlin~\cite{Haase2019}.
After this announcement, studies investigating how developers were dealing with the adoption of Kotlin as an official language concluded that developers believe that the use of Kotlin can improve the code quality, readability, and productivity~\citep{migration2020, Oliveira2020}. 

Kotlin provides a different approach to write applications because it combines object-oriented and functional features, some of them not present in Java or not available for Android development~\citep{KotlinComparison2016}.
Despite the increasing use of Kotlin among Android developers, to the best of our knowledge, there is no study in the literature about the adoption of Kotlin features by Android developers.
As pointed by \citet{Mazinanian:2017:UUL:3152284.3133909}, the lack of this knowledge negatively affects four audiences:
\begin{inparaenum}[\it i)]
\item researchers are not aware of the research gaps 
(i.e., the actual unsolved problems faced by the developers)
and thus miss opportunities to improve the current state of the art,
\item language and library designers do not know if the developers effectively use the programming constructs and APIs they provide are effectively used by the developers, or are rather misused or underused,
\item tool builders do not know how to tailor their tools, such as recommendation systems and code assistants, to the developers' actual needs and practices when using Kotlin,
\item developers are not aware of the good and bad practices related to the use of Kotlin features.
\end{inparaenum}

For those reasons, the goal of this paper is to understand the usage of Kotlin features on Android applications. 
In order to explore the usage of Kotlin features, we considered four aspects of features usage:
\begin{inparaenum}[\it i)]
\item {\bf which} features are adopted,
\item {\bf what} is the degree of adoption,
\item {\bf when} are these features added into Android applications for the first time, and 
\item {\bf how} the usage of features evolves along with applications' evolution.
\end{inparaenum}

To carry out our experiment, we used the largest publicly available dataset with open-source Android applications written in Kotlin~\citep{GoisMateus2019}, which has 387 applications with Kotlin code.
To study the adoption of Kotlin features, we extracted features from the source code of applications, and we identified when features were used for the first time.
To understand how the use of features evolves,
we analyzed the code repository of each application (i.e., Git) to mine the features used on each version (commit).
Then, we automatically assigned to each application-feature  pair $(a,f)$ an evolution trend that better describes the use of $f$ along with the evolution of $a$.
Different from other studies that focus on manual classification of evolution trends \citep{Hecht2015,Malavolta2018}, our method is completely automated.
Finally, we computed the most frequent evolution trend associated with a feature.


\newcommand{\rqone}{Which Kotlin features are adopted by Android developers?}  

\newcommand{\rqtwo}{When do Android developers introduce Kotlin features during applications' evolution? 
}

\newcommand{\rqtwob}{
What is the order of addition of Kotlin features and in what proportion these features are added together?
}

\newcommand{\rqthree}{How the usage of Kotlin features evolves along with the evolution of Android applications?}

As result of our research, this paper makes the following contributions:

\begin{itemize}
\item A study about the adoption of Kotlin features in the development of Android applications.
\item A methodology to automatically identify and classify trends of feature adoption during application's evolution.
\item  A study that shows how Kotlin features evolve during the applications' evolution.
\item A list of 7 main empirical findings about the usage and evolution of Kotlin and its implications to developers, researchers, tool builders.
\end{itemize}

All the data presented in this paper is publicly available in our appendix: \url{https://github.com/UPHF/kotlin_features}


\section{Description of Kotlin's Features}
\label{sec:feautres}

\subsection{What is Kotlin?}
\label{sec:kotlinintro}
Kotlin is a programming language that runs on the Java virtual machine. 
It combines object-oriented and functional features and, as it has 100\% interoperability with Java.
Therefore, developers can use it to write new files in an existing Java project or to write a new application using Kotlin from scratch.
Moreover, the official IDE for Android development, Android Studio, provides first-class support for Kotlin, including a built-in tool to convert Java-based code to Kotlin. 
Android applications are historically developed using Java and, by default, the Android platform is compatible with Java 6, which means that several new Java features, such as \emph{Lambdas}, cannot be used for developing Android apps.
Using Android Studio 3.0 or later, developers can use any feature from Java 7 and some features from Java 8, such as \emph{Lambdas}.  
However, that requires to use a minimum target Android SDK version~\citep{AndroidDevelopers2017}.
Therefore, the announcement made by Google brought the possibility for developers to use Kotlin instead of Java to write code for Android and, consequently, to use Kotlin features not supported by Java.

\subsection{Selection of Kotlin features}
\label{sec:method:selection}

When a new programming language is released, it offers developers a set of language \emph{features}. Other languages could already provide some of these features, whereas other features can be completely new.
In this paper, we exclusively focus on features that are available in Kotlin but not in Java.
Our goal is to study how Android developers use programming features that were not fully available for developing Android applications before the release of Kotlin.
Despite the focus of this study, we claim that the methodology presented could be applied to investigate the usage of Kotlin features in a context other than mobile development (e.g., server-side applications).

To identify Kotlin features not available in Java, we inspected the Kotlin official website.
First, we extracted 13 features from a document that compares Kotlin and Java~\citep{KotlinComparison2016}.
Then, we extracted 4 features from Kotlin's release notes, which were not mentioned in the comparison document (\textit{coroutine} as experimental feature and \textit{type alias} from release 1.1 and \textit{contract} and \textit{inline class} from release 1.3). 
Finally, we passed over the Kotlin Reference~\citep{kotlin_doc} and identified 7 more features.
Table~\ref{tab:feature:list} summarizes the features that we target in this study.

\newcounter{fid}
\setcounter{fid}{1}

\begin{table}[ht]
\centering
\scriptsize
\caption{Kotlin features and their release version. 
}
\begin{tabular}{|c p{3.5cm} p{1cm} p{2.5cm}|}
\hline
\textbf{ID} & \textbf{Feature} & \textbf{Release version} & \textbf{Normalization Criteria} \\\hline
\thefid{}\stepcounter{fid} & Type inference & 1.0 & \# of variable declarations \\\hline
\thefid{}\stepcounter{fid} & Lambda & 1.0 & LLOC \\\hline
\thefid{}\stepcounter{fid} & Inline function & 1.0 &  \# of named functions \\\hline
\thefid{}\stepcounter{fid} & Null-safety (Safe and Unsafe calls) & 1.0 & LLOC \\\hline
\thefid{}\stepcounter{fid} & When expressions & 1.0 & LLOC \\\hline
\thefid{}\stepcounter{fid} & Function w/arguments with a default value & 1.0 & \# of functions + \# of constructors \\\hline
\thefid{}\stepcounter{fid} & Function w/ named arguments & 1.0 & \# of function calls \\
\thefid{}\stepcounter{fid} & Smart casts & 1.0 & LLOC\\\hline
\thefid{}\stepcounter{fid} & Data classes & 1.0 & \# of classes \\\hline
\thefid{}\stepcounter{fid} & Range expressions & 1.0 & LLOC \\\hline
\thefid{}\stepcounter{fid} & Extension Functions & 1.0  & \# of named functions\\\hline
\thefid{}\stepcounter{fid} & String template & 1.0 & \# of strings\\\hline
&&&\# of properties\\\cline{4-4}
\multirow{-2}{*}{\thefid{}\stepcounter{fid}} &
\multirow{-2}{*}{Delegation (Super and Property)} &
\multirow{-2}{*}{1.0}
&\# of inheritances \\\hline
\thefid{}\stepcounter{fid} & Operator Overloading & 1.0 & \# of named functions\\\hline
\thefid{}\stepcounter{fid} & Singleton & 1.0 & \# of object declarations\\\hline
\thefid{}\stepcounter{fid} & Companion Object&  1.0 & \# of object declarations \\\hline
\thefid{}\stepcounter{fid} & Destructuring Declaration & 1.0 & \# of variable declaration \\\hline
\thefid{}\stepcounter{fid} & Infix function & 1.0 & \# of named functions \\\hline
\thefid{}\stepcounter{fid} & Tail-recursive function & 1.0  & \# of named functions\\\hline
\thefid{}\stepcounter{fid} & Sealed class & 1.0 & \# of classes\\\hline
\thefid{}\stepcounter{fid} & Type aliases & 1.1 & LLOC\\\hline
\thefid{}\stepcounter{fid} & Coroutine (experimental) & 1.1 & LLOC \\\hline
\thefid{}\stepcounter{fid} & Contract (experimental) & 1.3  & LLOC\\\hline
\thefid{}\stepcounter{fid} & Inline class & 1.3 & \# of classes\\
\hline
\end{tabular}
\label{tab:feature:list}
\end{table}

\section{Methodology}
\label{sec:methodology}


The goal of this paper is to understand the usage of Kotlin features on Android applications.
The following research questions guide our study:

\begin{itemize}
    \item $RQ_1$: \rqone
    \item $RQ_2$: \rqtwo
    \item $RQ_3$: \rqthree
\end{itemize}

In this section, we present the study design applied to respond to these research questions.
First, we present a method to detect Kotlin features from source code (Section~\ref{sec:method:tooldetect}).
Second, we present a method to identify when these features are used for the first time within an application (Section~\ref{sec:miningapps}).
Finally, we present a method to classify the use of features along with the application's evolution (Section~\ref{sec:met:trend}).

\subsection{Identification of Kotlin features}
\label{sec:method:tooldetect}

For responding to our research questions,
we need to identify the use of Kotlin features from the applications' source code.
For that reason, we built a tool that given as input a Kotlin source code file (.kt file) and returns a list of all features found in that file.

We built the tool as follows.
Our feature detection tool operates on the abstract syntax tree (AST) provided by the Kotlin compiler API. 
For each feature presented in Table \ref{tab:feature:list},
we first manually investigated how a feature is represented on an AST.
Then, we encoded different analyzers for detecting feature instances on ASTs. 
We encoded analyzers successfully for the 24 features presented in Table~\ref{tab:feature:list}.
We built 26 analyzers because we encoded two analyzers for two features: \rfeat{Null-safety} and \rfeat{Delegation}. 
We split the \rfeat{Null-safety} feature in two: 
\begin{inparaenum}[\it 1)]
\item \rfeat{Safe call} that provides information about the usage of the safe call operator `?' and
\item \rfeat{Unsafe call} that tells whether developers use the not-null assertion operator `!!' that we will refer as unsafe operator. 
\end{inparaenum}
We also split \rfeat{Delegation} in two features: \begin{inparaenum}[\it 1)]
\item \rfeat{Super Delegation} and 
\item \rfeat{Property Delegation}.
\end{inparaenum}
Moreover, regarding the feature \textit{Type inference}, our analyzer focuses on a single scenario: variable declaration (e.g. ``var a=10;'', the type of $a$ is inferred (int)).

To the best of our knowledge, there is no benchmark of Kotlin features usage that we could use for evaluating our tool.
Therefore, we run an experiment based on manual verification to evaluate its precision.
We executed our tool over the last version of each application from our dataset (387 applications), which returned a list of features instances found, with their respective locations (file name and line number).
This information allowed us to verify whether each reported feature instance was present or not at the reported files.
To achieve a confidence level of 95\% and a confidence interval of 10\% we checked 96 instances of each feature, randomly selected. 
The evaluation results, available on our appendix, showed the precision of the tool is 100\%.\footnote{To find the minimum number of instances to analyze (96), we compute the confidential level by considering as sample size the number of instances of the most frequent feature, which was \textit{type inference} with \numprint{165667} instances.
}
To measure the recall of our tool, both authors manually analyzed 100 files randomly selected. 
Then, we executed our tool over this set of files and calculated the recall.
we found a recall of 100\% for all features, but \textit{coroutine} (an experimental feature). Our strategy based on keywords could not identify all possible \textit{coroutines}, resulting in a recall of 91\%.

\subsection{Mining Kotlin Features from Applications}
\label{sec:miningapps}

\subsubsection{Analyzing Feature Evolution by Inspecting Commits}
\label{sec:inspectcommits}

To analyze the usage of features along with the history of one application,
we created another tool that takes as input a Git repository and produces, for each version $v$ (i.e., commit) the number of features found on $v$. 
The tool navigates through the commits of the active branch, in general, the \textit{master} branch. 
Given a Git repository, it starts from the oldest commit, and for each commit, it computes the number of features by invoking our feature detection tool described in Section \ref{sec:method:tooldetect}. 
When a repository is analyzed, our tool generates a JSON file. This file has for each commit, the number of features of each studied feature grouped by file.
This tool is built over Coming~\citep{coming2019}, a framework for navigating Git repositories that allow users to plug-in their source code analyzers. 

\subsubsection{Summarizing use of Kotlin features}
\label{sec:feature_intro}

To answer the first research question (RQ$_1$), we processed the output of our feature evolution tool (Section \ref{sec:inspectcommits}). 
For each feature $f$, we counted the number of applications that have at least one instance of $f$ in any commit and
the total number of \emph{instances} in the last commit of every application. 

\subsubsection{Normalization of feature instances}

As the applications may have different sizes, we normalized the number of instances following the criterion presented in Table~\ref{tab:feature:list}. 
We could normalize each feature with a unique metric of size, such as LLOC.
However, we consider that it would be more meaningful (and would better describe the use of a feature) if we normalize each feature $f$ by a metric that is related to $f$.
For instance, it gives more information to say that an application has 1 data class each $N$ classes, rather than reporting that it has 1 data class every $Y$ lines of code.

Now, we detail the normalization process.
As explained in Section~\ref{sec:method:tooldetect}, our analyzer of \textit{type inference} considers only variable declarations. 
Thus, we normalized the instances of this feature by the number of variables declared.
Since \textit{destructuring declarations} break down objects into (declaring) multiple variables, we also normalized its instances by the number of variables declared.

In Kotlin, every function declared is a \textit{named function} node in the AST. 
Thus, we normalized the number of instances of \textit{extensions functions} by the number of \textit{named functions}.
Since only \rfeat{named functions} might receive the modifier \rfeat{inline}, we used their number to normalize the number of \textit{inline functions}. 
For the same reason, we used the same criterion to normalize the number of \textit{tail-recursive}.
Analogously, we normalized the number of \textit{data classes}, \textit{sealed classes} and \textit{inline class} by the number of classes.
Additionally, the number of \textit{named functions} was also used to normalize the number of \textit{operator overloading} because, by definition, an overloaded operator is a \textit{named function} that receives the modifier \textit{operator}.

Arguments with a default value can be used in \textit{named functions} and \textit{constructors}.
Thus, we normalized \textit{function with arguments with a default value} by the number of \textit{named functions} and \textit{constructors}. \textit{Named arguments} are used when a method/function is called. Consequently, we normalized the number of function calls with named arguments by the number of function calls. 
Besides, since only strings might have a \textit{string template}, we normalized the number of \textit{string templates} by the number of strings. Concerning Kotlin delegations, we normalized the number of properties delegated by the number of properties.
Moreover, as \textit{super delegation} is an alternative to inheritance, we normalized their instances by the number of classes. In Kotlin, object expressions are used to declare singletons and companion object.
Consequently, we normalized the number of \textit{singleton} and \textit{companion object} by the number of object declaration.
For the remaining features, we normalized them by LLOC because we could not find a better criterion.

\subsubsection{Identifying the first use of Kotlin features}
\label{sec:method:timeintro}

For responding to RQ$_2$, we computed, for each application $a$ and for each feature $f$, the first commit $C_{fa}$ that introduces an instance of $f$ into $a$.  
Finally, we defined a metric, named \textit{introduction moment}, $m_{af} \, \in \, [0, n]$ where $n$ is the number of days between the first Kotlin commit and the last commit, that measures how long after the initial commit a feature $f$ was introduced into $a$. It is expressed in days.
For instance, $m_{af} \, = \, 0$ means that feature $f$ was introduce into $a$ in the same day of the first Kotlin commit,  $m_{af} \, = \, 5$  means that $f$ was added in the 5 days after the Kotlin introduction, and $m_{af} \, = \, n$, means that $f$ was introduced in the same day of the last commit.

\subsection{Classifying the use of features along with the application's evolution}
\label{sec:met:trend}

The goal of $RQ_3$ is to detect trends that describe the usage evolution of features along with the applications' history.
For example,
we want to detect applications where the use of a particular feature is constant, increases or decreases along with its evolution.

\subsubsection{Feature evolution on applications}
To study the use of a feature $f$ along with the history of one application, we first counted the number of \emph{instances} of $f$ on each applications' version.

\subsubsection{Classification of feature evolution trend}
\label{sec:classification}
To classify each pair of application-feature $(a,f)$ and find the trend that \emph{best} describes the evolution of a feature $f$ in the history of an application $a$, we defined a set of evolution trends following a data-driven approach.
First, we automatically plotted a two-dimension plot for each application-feature $(a,f)$ where the axis X corresponded to the number of commits (chronologically order) from $a$, and the axis Y was the number of instances of a feature $f$. 
Then, we iteratively analyzed the generated plots.
For each of them, we first described the trend we observed in two ways: 
\begin{inparaenum}[\it 1)]
\item natural language description (e.g., ``Instances increase''), 
\item mathematically (e.g., ``y=ax+b'').
\end{inparaenum}
Then, we checked whether a similar trend had been seen before based on such descriptions. 
We repeated this last step until no new trend be found after analyzing dozens of plots.
To avoid bias, both authors have analyzed the same plots separately, and later, we compared and discussed the result found.
We found 11 unique evolution trends that were described by 6 different mathematical formulas. 

Other works have previously defined evolution trends, such as \citet{Hecht2015} and \citet{Malavolta2018}.
Both studies have used their trends to classify \emph{manually} evolution plots.
On the contrary, the definition of our evolution trends was motivated by the need to classify trends automatically, which guarantees
\begin{inparaenum}[\it i)]
\item  scalability (i.e., thousands of apps-feature pairs to classify) and 
\item the replication (analysis of other applications) of this study.
\end{inparaenum}

\subsubsection{Considered trends and formulas}
\label{sec:met:trends}
\begin{figure*}
    \centering
    \captionsetup[sub]{font=scriptsize, justification=centering}
    \begin{subfigure}[t]{0.085\textwidth}
        \scalebox{-1}[1]{\includegraphics[width=\textwidth]{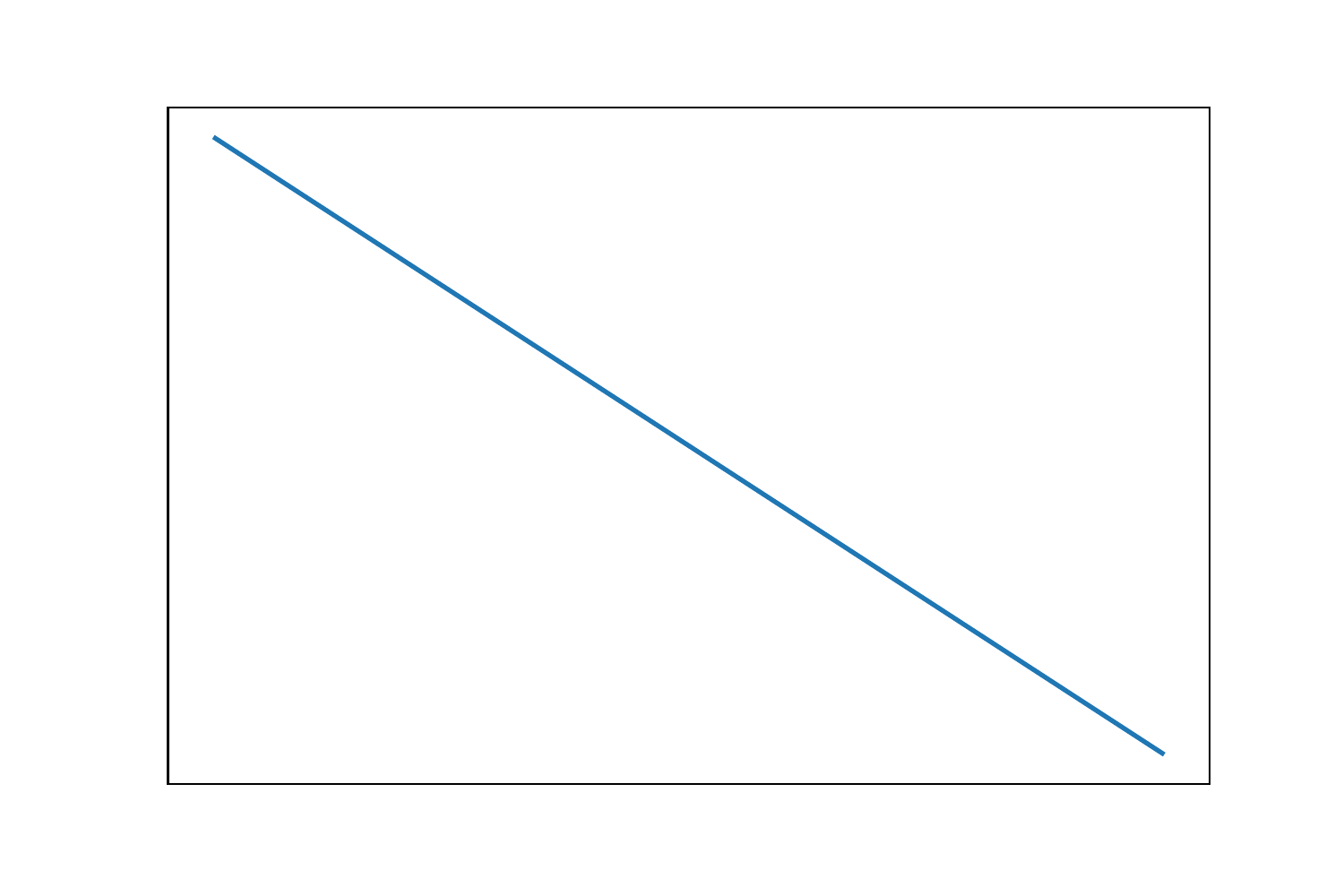}}
    \caption{Constant \\ Rise}
    \label{fig:sample:T1}
    \end{subfigure}
    \begin{subfigure}[t]{0.085\textwidth}
        \includegraphics[width=\textwidth]{t2_sample.pdf}
    \caption{Constant \\ Decline}
    \label{fig:sample:T2}
    \end{subfigure}
    \begin{subfigure}[t]{0.085\textwidth}
        \includegraphics[width=\textwidth]{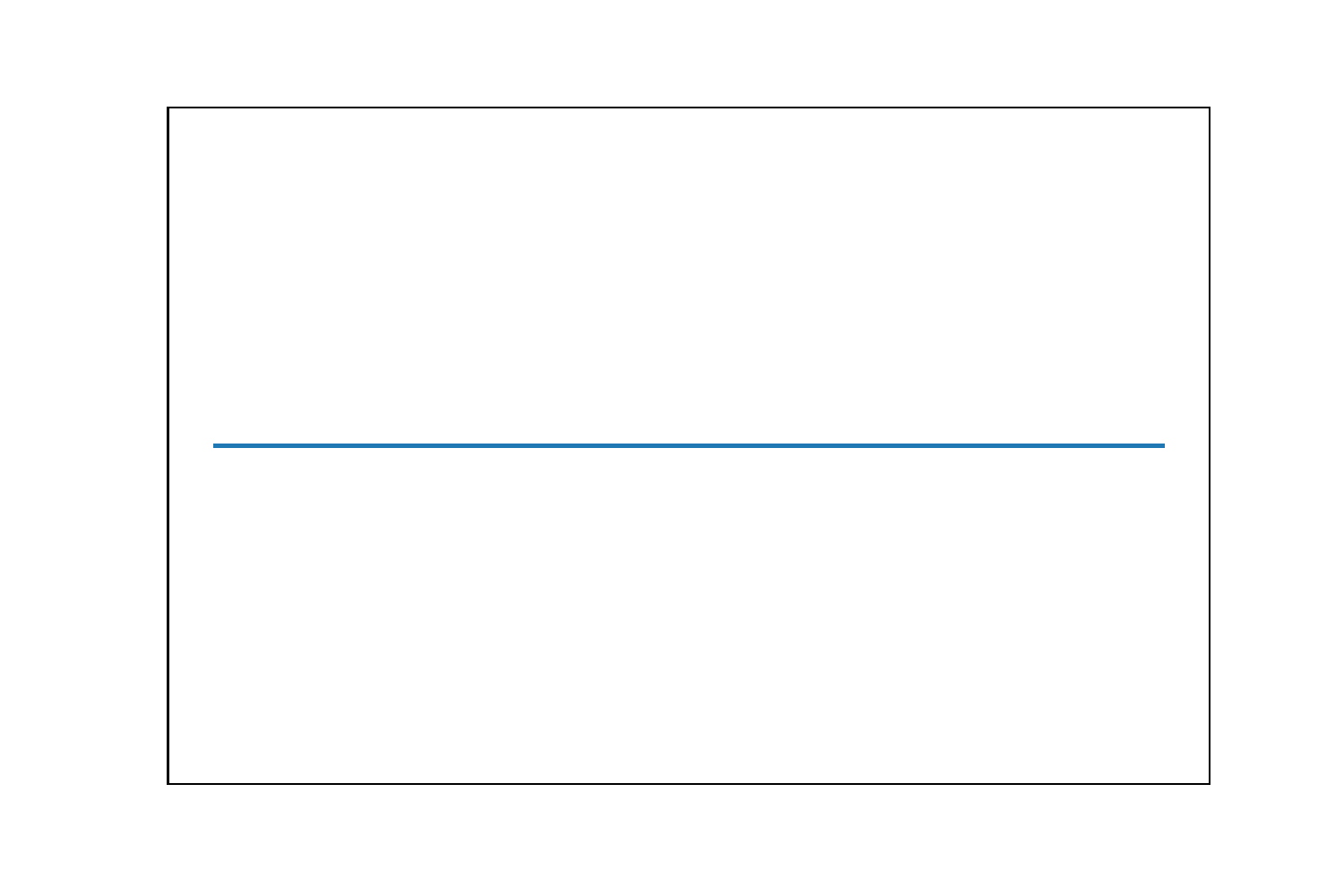}
    \caption{\etthree}
    \label{fig:sample:T3}
    \end{subfigure}
    \begin{subfigure}[t]{0.085\textwidth}
        \scalebox{-1}[1]{\includegraphics[width=\textwidth]{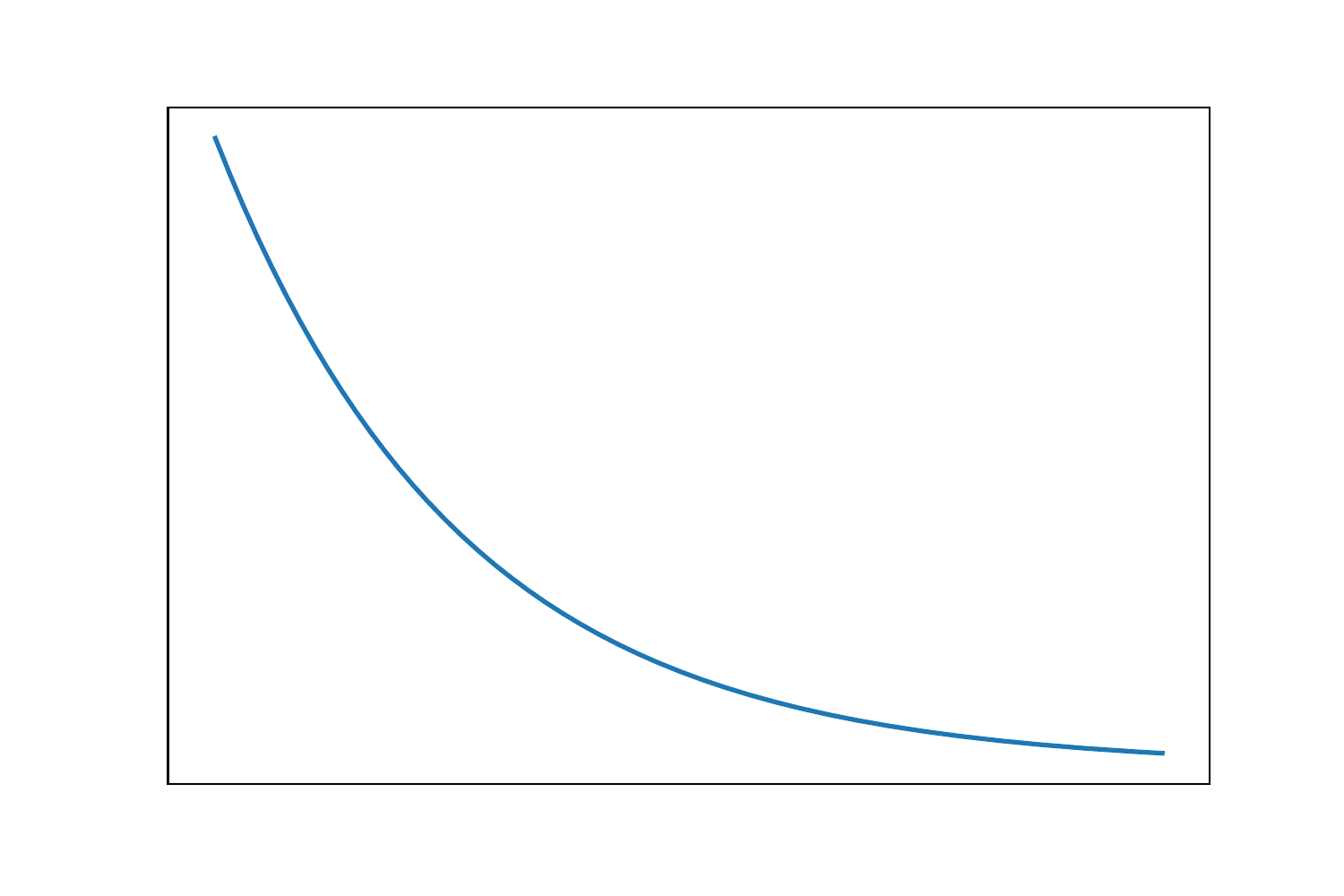}}
    \caption{\etfour}
    \label{fig:sample:T4}
    \end{subfigure}
    \begin{subfigure}[t]{0.085\textwidth}
        \includegraphics[width=\textwidth]{t5_sample.pdf}
    \caption{Sudden \\ Decline}
    \label{fig:sample:T5}
    \end{subfigure}
    \begin{subfigure}[t]{0.085\textwidth}
        \includegraphics[width=\textwidth]{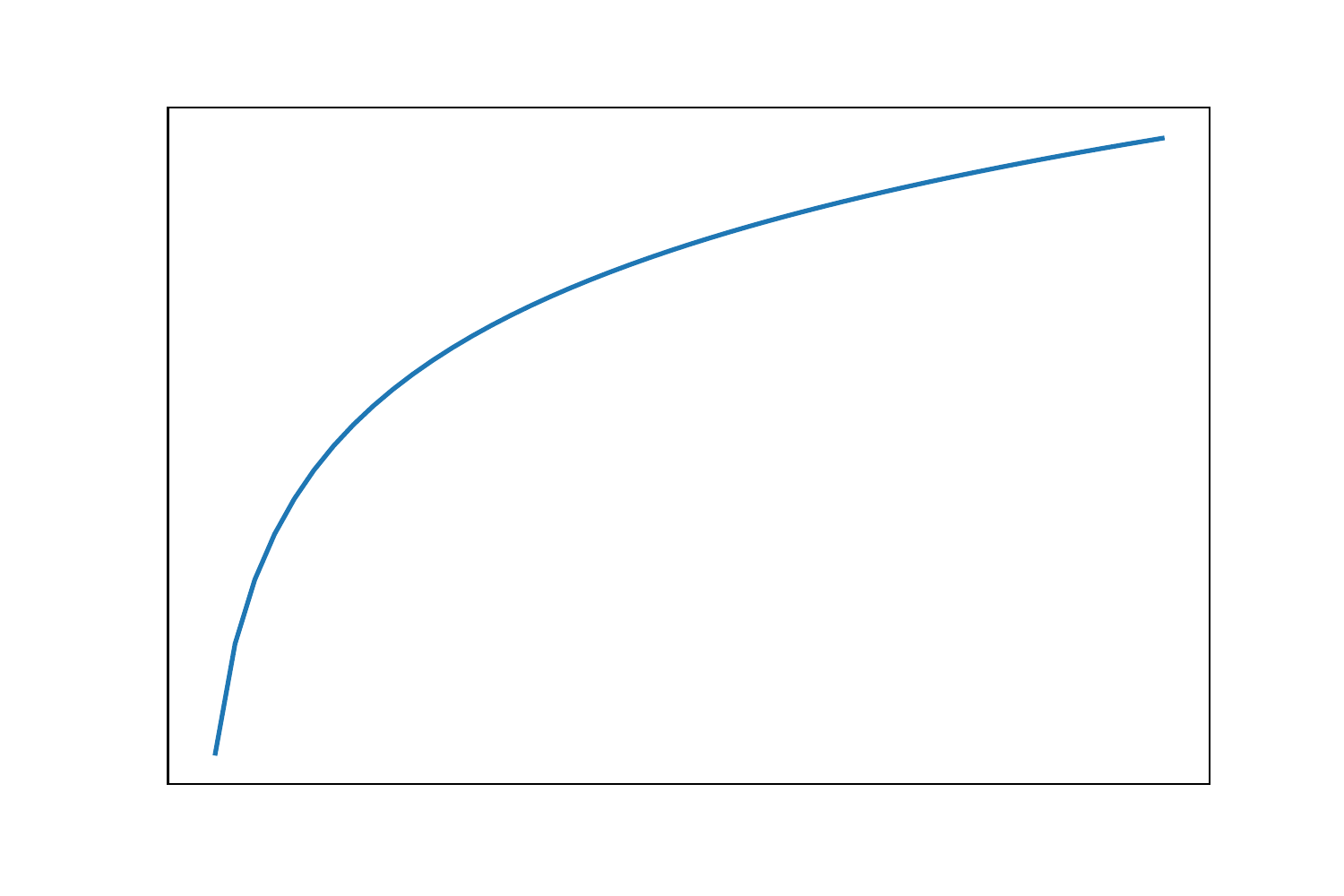}
    \caption{Sudden Rise \\ Plateau}
    \label{fig:sample:T6}
    \end{subfigure}
    \begin{subfigure}[t]{0.085\textwidth}
        \includegraphics[width=\textwidth]{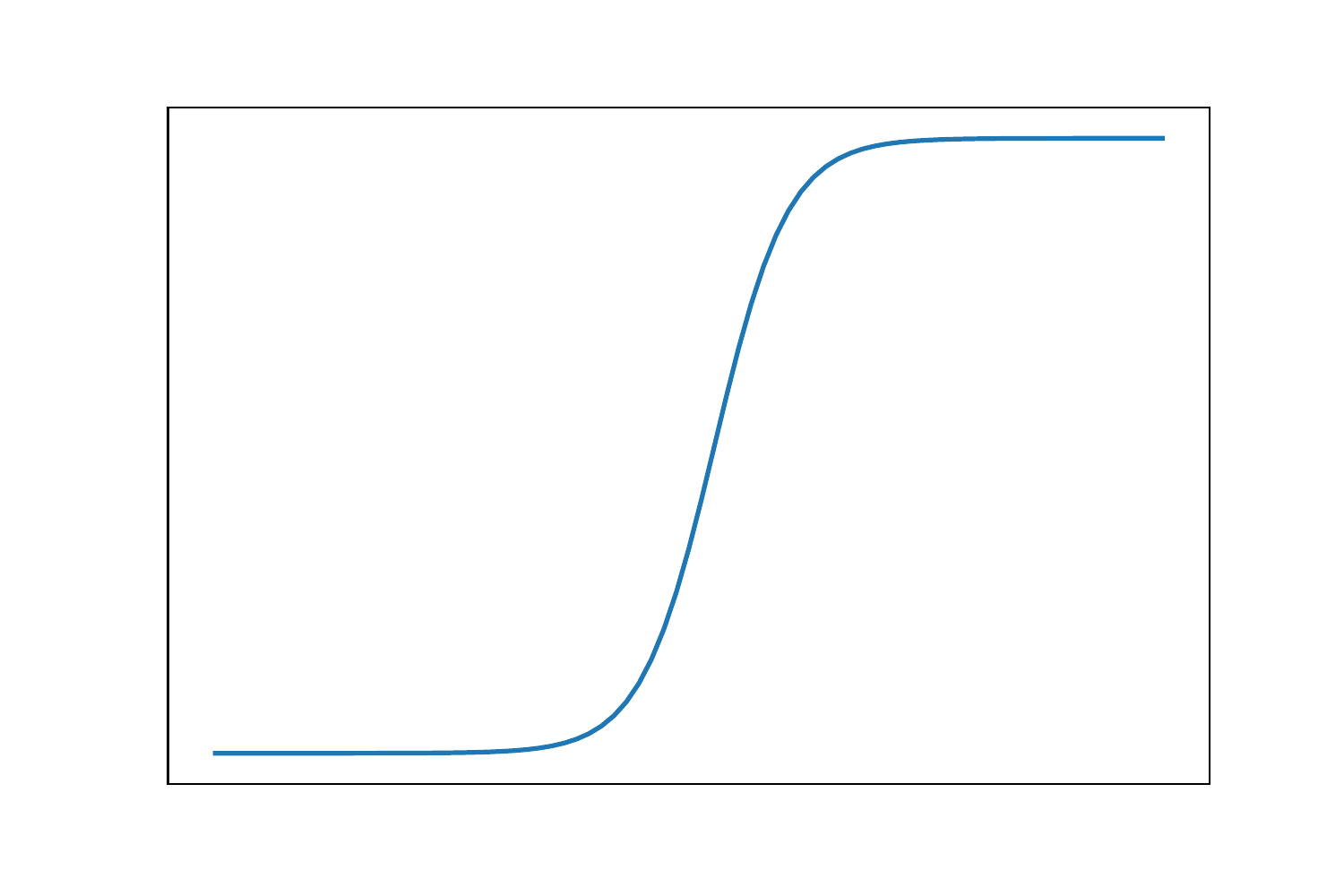}
    \caption{\etseven}
    \label{fig:sample:T7}
    \end{subfigure}
    \begin{subfigure}[t]{0.085\textwidth}
        \scalebox{-1}[1]{\includegraphics[width=\textwidth]{t7_sample.pdf}}
    \caption{Plateau Gradual \\ Decline}
    \label{fig:sample:T8}
    \end{subfigure}
    \begin{subfigure}[t]{0.085\textwidth}
        \includegraphics[width=\textwidth]{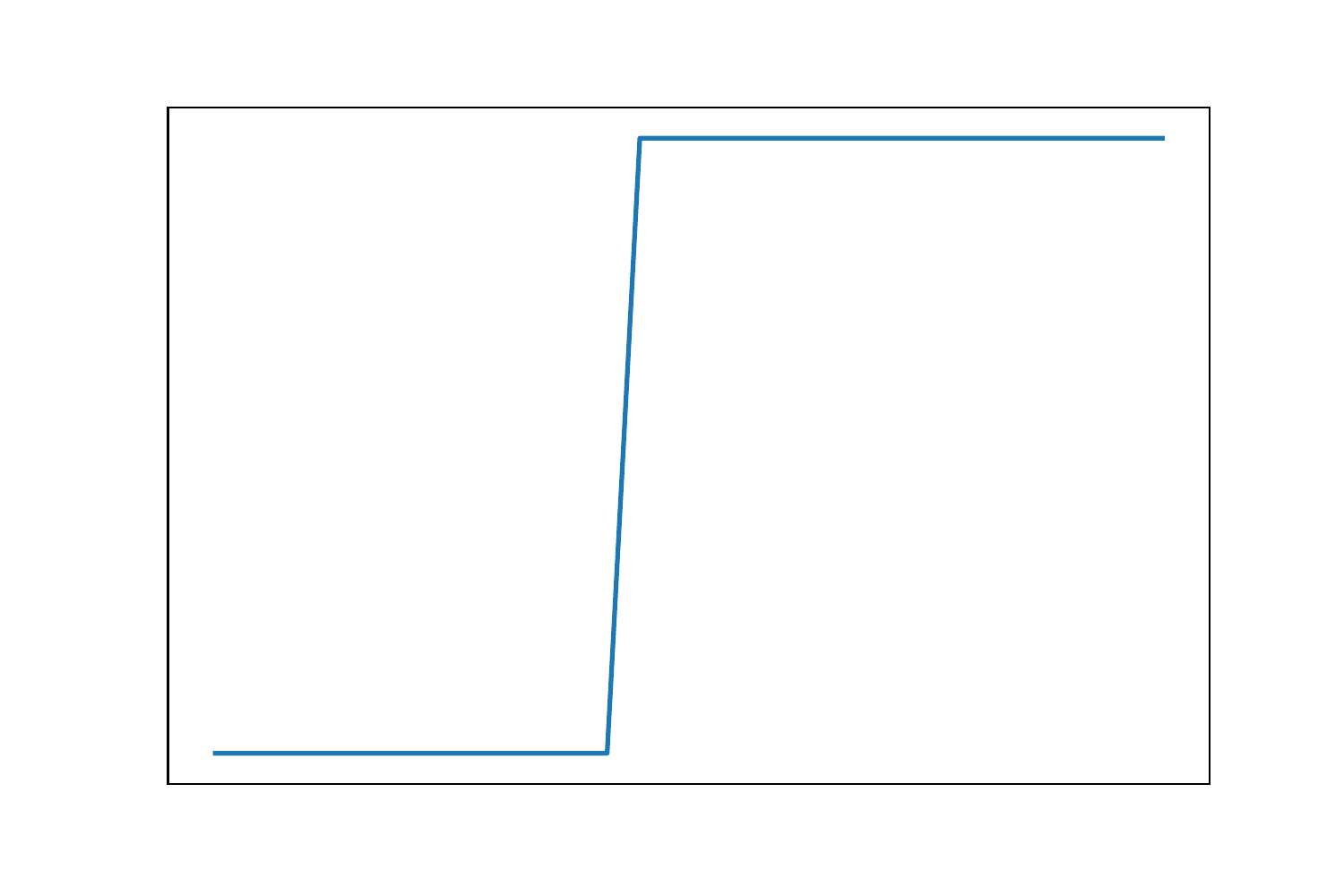}
    \caption{Plateau Sudden \\ Rise}
    \label{fig:sample:T9}
    \end{subfigure}
    \begin{subfigure}[t]{0.085\textwidth}
        \includegraphics[width=\textwidth]{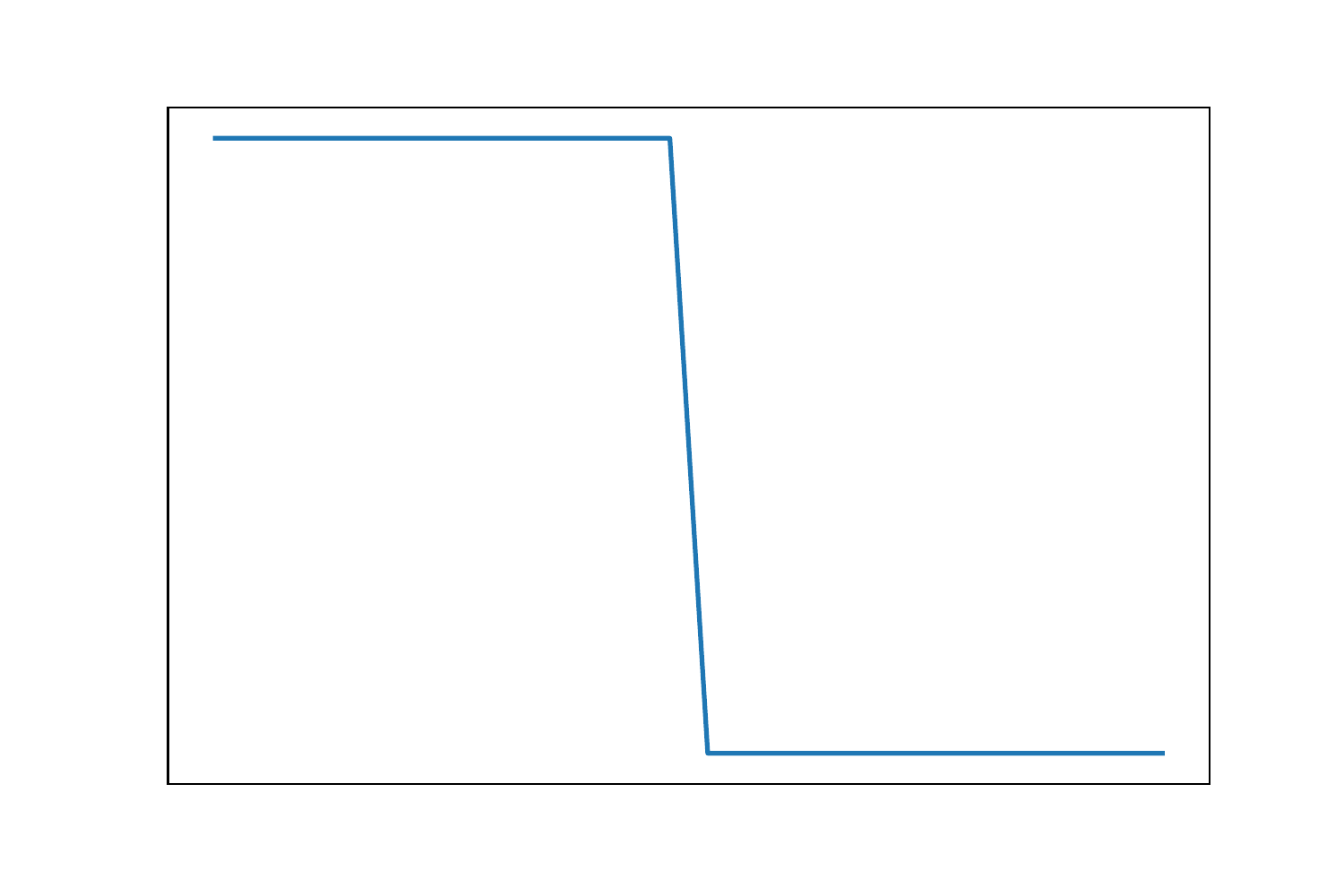}
    \caption{Plateau Sudden \\ Decline}
    \label{fig:sample:T10}
    \end{subfigure}
    \begin{subfigure}[t]{0.085\textwidth}
        \includegraphics[width=\textwidth]{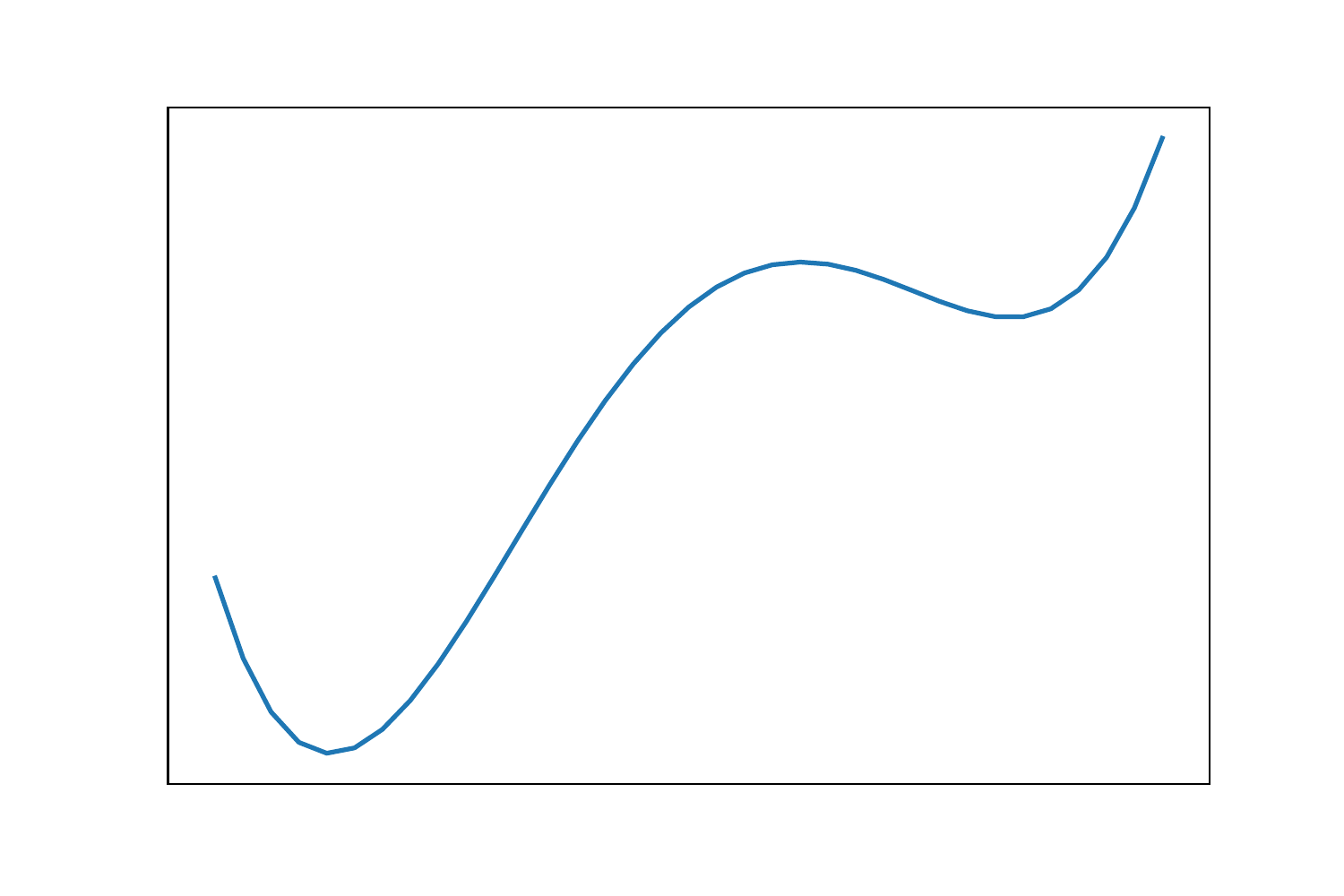}
    \caption{\eteleven}
    \label{fig:sample:T11}
    \end{subfigure}
    
\caption{Example of evolution trends. The x-axis shows the evolution of an application, i.e., commits, and the y-axis shows the number of occurrences of a feature.}
\label{fig:trend:sample}   
\end{figure*}

We established 11 major feature evolution trends to answer RQ3.
An example of each of them is shown in Figure~\ref{fig:trend:sample}. These trends are:

\textit{\etone~(\etoneabb)} describes features that once they are introduced (i.e., used for first time in an application), developers tend to add more instances of this feature in future application versions. 
Therefore, the number of instances of increases at a constant rate, i.e., linearly along with the application's evolution.

\textit{\ettwo~(\ettwoabb)} describes features that once they are introduced, developers tend to remove them gradually in future applications' versions. 
Therefore, along with the application's evolution, the number of feature instances decreases at a constant rate. 

\textit{\etthree~(\etthreeabb)} describes features whose numbers of instance remain the same after its introduction, along the application's evolution.

We used the linear function given by the formula $ y = a x + b$ to detect \etoneabb, \ettwoabb~and \etthreeabb.
Since in linear function, the rate of change (given by coefficient $a$) is always constant,
we could classify the application's trend into:
\begin{inparaenum}[a)]
\item[] (\etoneabb) when $a > 0$, which implies on constant increase;
\item[] (\ettwoabb) when  $a < 0$, which implies on constant decrease; and
\item[] (\etthreeabb) when $a = 0$, which determines a constant behavior.
\end{inparaenum}

\textit{\etfour~(\etfourabb)} describes those features that the number of occurrences grows suddenly after relative stability along with the applications' history.
Using this trend, we are able to identify those features that present a small number of instances in the first commits and then, at the following commits, on each commit, developers introduce significantly more instances.

\textit{\etfive~(\etfiveabb)}, analogously, describes the opposite behavior of \etfourabb, where the number of feature instances decreases suddenly. 
Features that present many instances since the first commits and then, suddenly, start to be removed, and this behavior continues in the next consecutive commits.

We used the exponential function given by the formula $ y = ab^x + c$ to detect \etfourabb~and \etfiveabb.
In this formula, the rate of change is given by coefficient $b$.
When $b$ is greater than 1, the value of  $y$ increase as $x$ increases.
On the other hand, if $b$ is $0 < b < 1$, the value of a function increases as the value of $x$ decrease.
Then, considering the value of $b$ we classified evolution trends better described by an exponential function in two trends, \etfourabb~and \etfiveabb.

\textit{\etsix~(\etsixabb)} describes features which most of its instances are introduced in the firsts commits of an application. Then, during the rest of the application's history, only a few instances are introduced.

We used the logarithmic function given by the formula $ y = a\log_e(bx) + c$ to detect  \etsixabb. 
In this case, unlike the exponential formula, the rate of change always decreases with time. 

\textit{\etseven~(\etsevenabb)} describes those feature which once they are introduced, the number of instances tends to remain the same during an interval of commits and then presents a sudden increase in few consecutive commits, and finally presents a stable behavior again.

\textit{\eteight~(\eteightabb)} is similar to the trend \etsevenabb, this trend describes those features that the number of instances starts and finishes stable. 
However, on the contrary of \etsevenabb, here, the change that happens between the periods of stability is a reduction of the number of instances in a few consecutive commits.

\textit{\etnine~(\etnineabb)} is similar to \etsevenabb. However, using this trend, we aim to detect features that present:
\begin{inparaenum}[a)]
\item a stability period at the beginning of the application's evolution
\item a transition period containing only one commit and
\item stability period at the end of the application's evolution.
\end{inparaenum}

\textit{\etten~(\ettenabb)} analogously to \etnineabb, this trend is a special case of \eteightabb~where the change between the two periods of stability has a marked decrease in the number of instances, in two consecutive commits.

We used the Sigmoid function, $y = \frac{L}{1 + \epsilon^(-k(x -x0))} + b$, to detect the trends \etsevenabb, \eteightabb, \etnineabb~and \ettenabb. 
A Sigmoid is a bounded function where $y$ can assume values from $-L + b$ to $L + b$.
Additionally, when $L * b > 0$, $y$ approaches $L +b $ as $x$ approaches $+\infty$ and approaches $-L + b$ as $x$ approaches $-\infty$.
On the other hand, we observe the opposite behavior, when $L * b < 0$.
While $x0$ is Sigmoid's midpoint, which marks the middle of the S-curve, $k$ determines the steepness of the curve.
The \etsevenabb~trend is characterized by the lowest the number of instances ($-L + b$) at the beginning of the application history and the highest value ($L + b$) of the number of instances in the end.
Moreover, the transition between the lowest and the highest values is characterized by a gradual rise, in a small number of commits, when developers introduce a considerable number of instances.
The \eteightabb~trend follows the opposite behavior, starting from the highest number of instances and finishing with  the lowest number of instances.
The differences between \etsevenabb~and~\etnineabb, and~\eteightabb~and~\ettenabb, is determined by the coefficient $k$, where
a small value of coefficient $k$ makes the slope of a Sigmoid function be completely vertical.

\textit{\eteleven~(\etelevenabb):} describes features that the number of instances alternates between an increase and a decrease during the application's evolution.

We used the polynomial function given by the formula $a_n x^n + a_{n-1}x^{n-1} + \dots + a_2 x^2 + a_1 x + a_0$, where $n$ is the degree of the polynomial and $a_n \, > 0$ to detect the \eteleven~trend.
Since the degree of a polynomial determines the number of minimum and maximum (at most $degree - 1$) given an open interval, using this formula we detected those features which the number of instances along the application's evolution form a curve with two or more minimum or maximum.

\subsubsection{Finding the formula that better describes the feature evolution}
\label{sec:met:fit}
To match a feature evolution trend from one application with one of the studied trends, we applied the following methodology:

\textit{Obtaining the series of the number of instances per application.}
As explained in Section~\ref{sec:miningapps}, given one application, our tool's execution gives each feature's number of instances that each application's version (i.e., commit) has.
Therefore, for each application $a$, we generated a series of values $y_{xa}=\{vax_1, vax_2, .., vax_n\}$ where $vax_i$ corresponds to the number of instance of feature $f$ detected in the $i$-version of $a$ (corresponding to the $i$-th commit considering chronological order, starting with the first version that introduces $x$). 
That is, the first element contains the number of features in the first version that introduces $f$. 

\textit{Computing formulas' coefficients by fitting the series.}
For each series $y_{xa}$ from a pair application-feature $(a,x)$, and for each formula $f$ presented in \ref{sec:met:trend}, we used non-linear least squares to fit function $f$ to data $y_{xa}$.
This gives as result a set of coefficients ($\alpha_1$, .., $\alpha_n$) for $f$ that correspond to the optimal values so that the sum of the squared residuals $SS_{reg}$ (Formula \ref{eq:ssres}) is minimized.

\begin{equation} 
\label{eq:ssres}
SS_{\rm res}=\sum_{i}^{|versions \, a|} (y_{xa_{i}} - f_i)
\end{equation}

The number of coefficients generated varies according to the formula $f$: for linear, the number of coefficients is two, whereas for a polynomial of 4 degrees is 5.
For executing the fitness of data, we used the function `curve\_fit' from library \textit{SciPy}~\citep{Scipy}.

\textit{Post-Processing formulas and coefficients.}
This step has two goals: 
\begin{inparaenum}[\it a)]
\item simplify polynomial formulas and
\item discard some Sigmoid functions which do not have a clear S-shape considering the domain applied $[0, \#commits - 1]$.
\end{inparaenum}
We simplify those polynomial formulas whose the coefficient $a_n$ is close to zero (i.e.,  $<$ than 0.0001), because these evolution trends can be similarly described by a polynomial $n-1$ degree.
We also discard Sigmoid functions where the coefficient $x_0$, i.e., the Sigmoid's midpoint, is outside the range of commits $[0, \#commits -1]$, because in these cases only one plateau could be in the commit range. 
Therefore, these cases do not configure any of the studied trends modeled with the Sigmoid function.

\textit{Choosing the formula that best represents a feature evolution  trend.} 
\label{sec:met:bestformula}
Once we computed the coefficients for each formula, we chose, for each pair $(a,x)$, the formula that yields less error to predict the values of  $y_{xa}$. We call it $f_{best\_ax}$.
Thus, this formula is considered as that one that best represents the feature evolution trend among all formulas we consider in this experiment (Section \ref{sec:met:trend}).
We based our choice on a statistical measure named R-squared (Coefficient of Determination) which measures how close the data are to the fitted formula. R-squared ($R^2$) is always between 0 (bad) and 1 (good).

When two formulas $f1$ and $f2$ produce the same $R^2$ values within a small positive delta (0.01 in our experiment), we prioritized
formulas with the smallest quantity of coefficients, that is:
$Linear >> Exponential >> Logarithmic >>  Sigmoid >> Polynomial$.

\textit{Summarizing of best formulas.}
For each feature $x$ and formula $f$, we count the number of applications that have $f$ as the best formula (Section \ref{sec:met:bestformula}).
The result from this step helps us to explain the most frequent evolution trend associated with each feature.

\subsection{Evaluation dataset}
\label{meth:dataset}

In this paper, we studied applications from FAMAZOA\footnote{\url{https://uphf.github.io/FAMAZOA/versions/v3}}~\citep{GoisMateus2019}, a dataset that has 387 open-source Android applications written fully or partially in Kotlin. 
It contains applications collected from 3 sources: AndroidTimeMachine~\citep{Geiger2018:data}, AndroZoo~\citep{Allix2016} and F-Droid\footnote{\url{http://f-droid.org}}
Those apps have a median of 6 contributors, 56 files, 3984 lines of Java/Kotlin and 64.08\% of them were updated in the last 6 months.

\section{Results}
\label{sec:results}


\subsection{RQ1: \rqone}
\label{sec:result:rq1}

\begin{figure}
    \centering
    \includegraphics[width=\columnwidth]{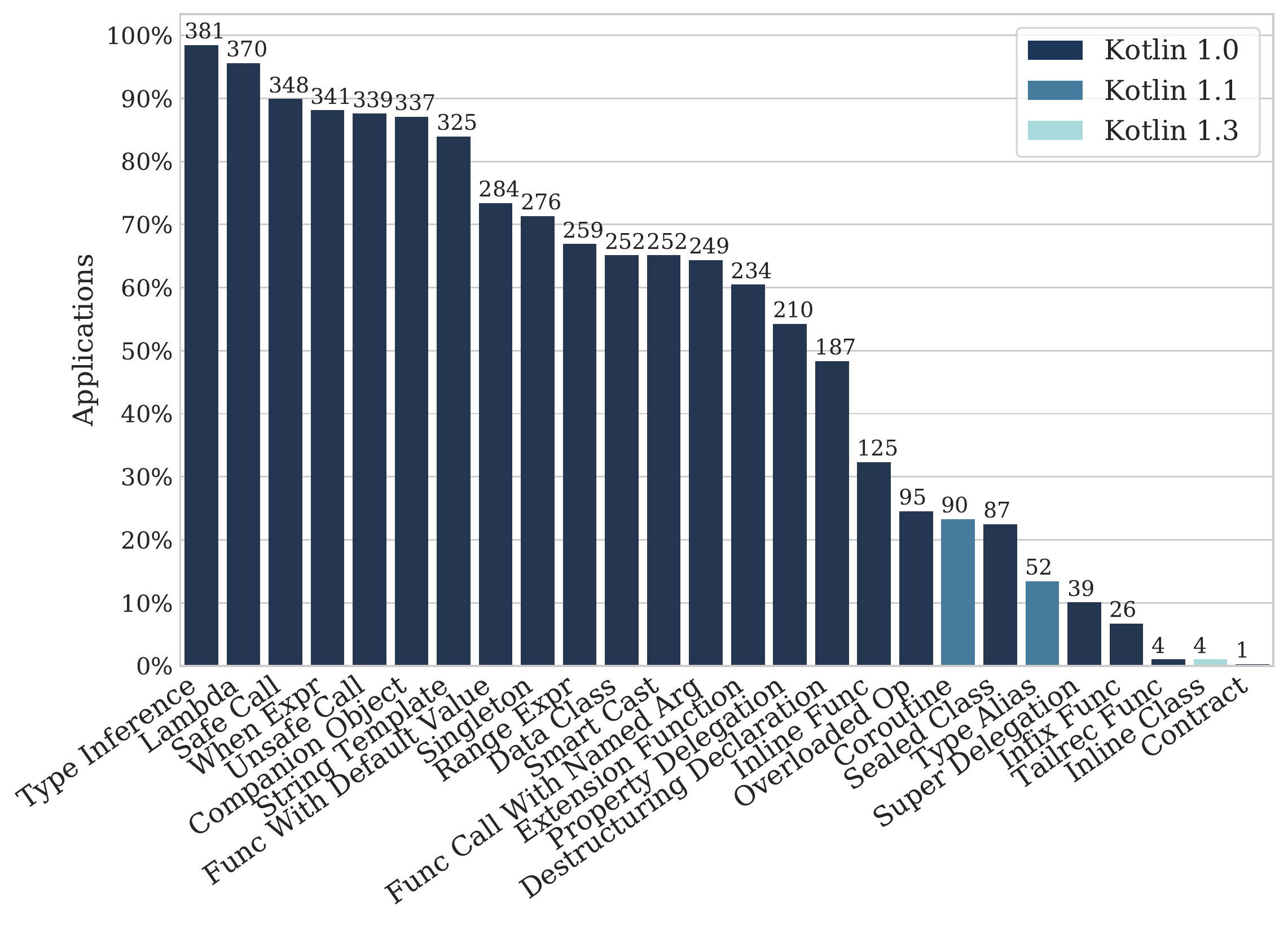}
    \caption{Percentage of applications that use a feature. Each bar corresponds to a feature and contains on top the number of applications that use that feature.}
    \label{fig:most_used_features}
\end{figure}

\begin{figure}
    \centering
    \includegraphics[width=\columnwidth]{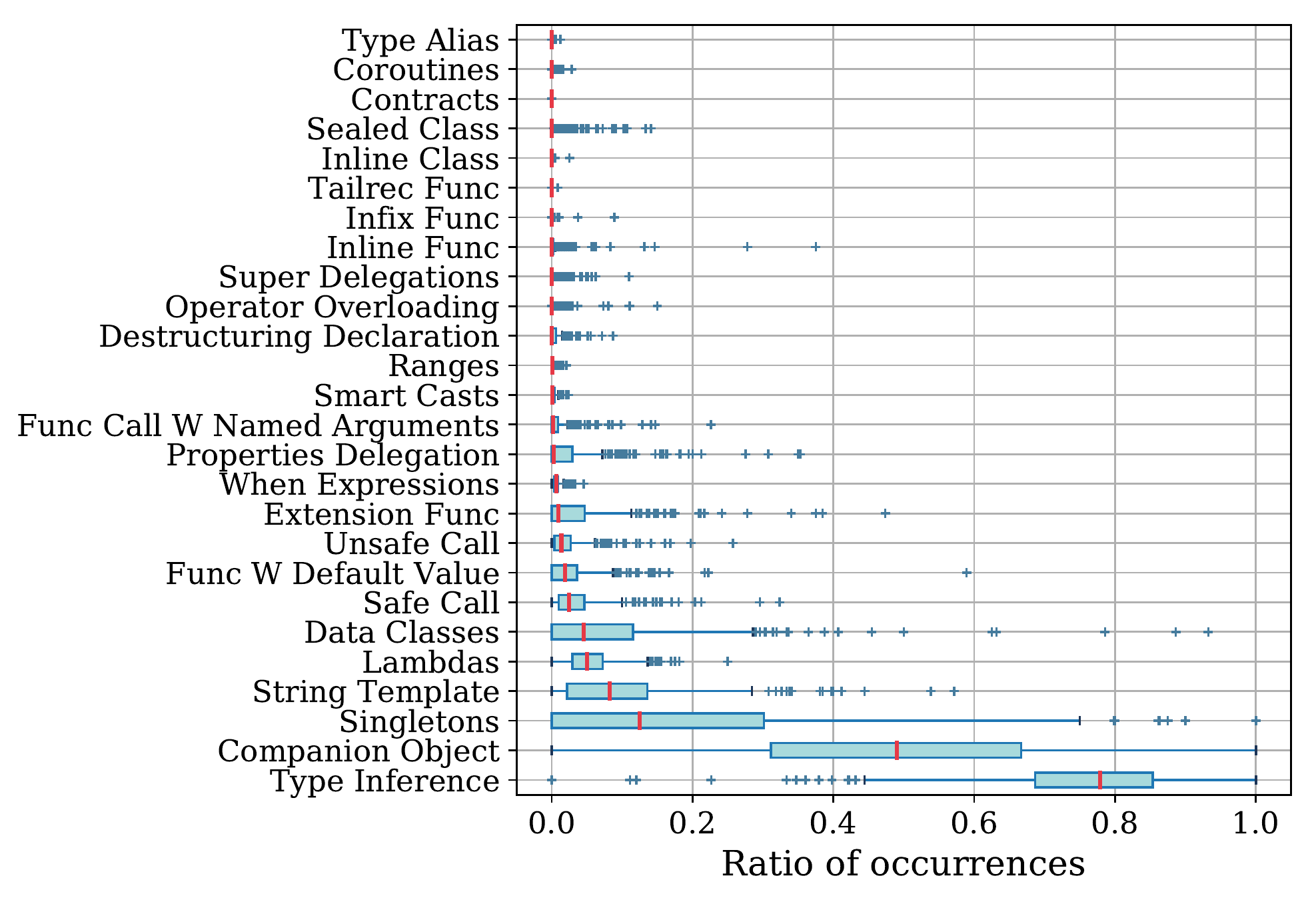}
    \caption{Kotlin features normalized.}
    \label{fig:box_normalized_features}
\end{figure}

To answer this research question, we applied the method described in Section~\ref{sec:miningapps}, and Figure~\ref{fig:most_used_features} summarizes its results.
For each feature, it shows the percentage of applications that have used that feature at least once (considering all the version of those apps). 
In addition, Figure~\ref{fig:box_normalized_features} shows the distribution of the normalized number of occurrences of each feature per application (considering the latest version of each one).

We observed that the most used feature is \rfeat{type inference} with 381 out of \datasetsize~(98\%) applications having at least one \rfeat{instance} of this feature, as Figure~\ref{fig:most_used_features} shows. 
We found that a median of 77\% of variable declarations do not have their type explicitly declared, as Figure~\ref{fig:box_normalized_features} displays. 
However, we noticed some applications whose all variables have their type inferred as well.
Therefore, we concluded that developers can write code more concisely, avoiding type declaration when the type is self-explained in the assignment.

\rfeat{Lambda} is the second most used feature, being found in 370 out of \datasetsize~(95\%) applications with a median of 71~\rfeat{instances} per application.
The fact that Android applications rely on callbacks to interact with the Android platform~\citep{yang2015static} could be seen as a reason for the number of instances found.
However, a high number of instances, especially when they are nested, could lead developers to write code with poor readability. 
In Kotlin, developers do not need to name the parameter of a single parameter lambda function. In this case, it is automatically named as `it'.
Therefore, a chain of nested lambdas using this mechanism might be hard to read. For that reason, JetBrains once considered removing the `it' parameter~\citep{KotlinDiscussions2017}.

\textit{Safe call} is the third most used feature. 
We found 348 out of \datasetsize~(89\%) applications where \rfeat{safe calls} were found, with a median of 28 occurrences per application.
Another feature related to null-safety, \textit{unsafe call}, is used in 339 out of \datasetsize~(87\%) applications.
As the opposite of \rfeat{safe calls}, the usage \rfeat{unsafe calls} could result in \textit{NullPointerException (NPE)}.
We investigated some instances of this feature manually, and we found cases where it could be substituted by a \textit{safe call} or the Elvis operator.

\begin{tcolorbox}[colback=white, boxrule=0.5pt, boxsep=0.0pt]
\textbf{Finding 1:}
   \textit{We found a median of 16 occurrences of unsafe calls per application, which makes these applications more prone to `NullPointerException'. 
    However, we found that occurrences could be replaced by Kotlin's built-in functions such as Elvis operator to avoid NPEs.
   }
   
\end{tcolorbox}

Considering \rfeat{companion objects}, the sixth most used feature, found in 337 out of 387~(87\%) applications, we observed that 48\% of the object declared are \rfeat{companion object}.
\textit{Companion objects} are the substitute of Java static members in Kotlin, but by default, properties and methods of \rfeat{companion objects} are not static. 
To make them static, developers should use the annotation \textit{@JvmField} or \textit{@JvmStatic}, and we found several not annotated objects in our study.

As Figure~\ref{fig:most_used_features} shows, \rfeat{function with arguments with a default value} and \rfeat{function calls with named arguments} are used in more than 60\% applications of our dataset with a median of occurrences of 3 and 2 occurrences per application respectively.
We normalized the number of \emph{instances} by the \emph{number of named function} plus \emph{number of constructors} and the \emph{number of function calls}, respectively, as Figure~\ref{fig:box_normalized_features} shows. 

\begin{tcolorbox}[colback=white, boxrule=0.5pt, boxsep=0.0pt]
\textbf{Finding 2:}
\textit{According to the Kotlin convention, the use of named arguments can improve the readability of the code~\citep{kotlin_conventions}. 
However, we found that less than 1\% of function calls have a named argument.
}
\end{tcolorbox}

Moreover, we found that 252 out of \datasetsize~(65\%) applications that use \rfeat{data classes}, with a median of 1 instance per application.
In total, we found that only the 4\% of applications' classes are data classes.
Figure~\ref{fig:box_normalized_features} shows the distribution of that proportion: for 25\% of applications, the proportion of data classes is more than 10\% of the total number of classes.

Kotlin provides simple approaches to use two well-known design patterns, \rfeat{Singleton} and \rfeat{Delegation}.
Regarding the usage of \rfeat{singleton}, we observed 276 out of \datasetsize~(71\%) applications have at least one class that implements Singleton pattern.
Furthermore, 
an application has a median 2 occurrence of this feature.

Considering \textit{properties delegation}, we observed it in 210 out of \datasetsize~(54\%) applications.
Normalizing the number of properties delegated by the number of properties for each application we found that: 

\begin{tcolorbox}[colback=white, boxrule=0.5pt, boxsep=0.0pt]
\textbf{Finding 3:} 
\textit{Although Kotlin standard library provides for several useful kinds of delegation, such as lazy and observable, less than 1\% of properties defined in Kotlin applications are delegated.}
\end{tcolorbox}

\rfeat{Coroutine} is most used feature considering those released after  Kotlin 1.0.
Note that \rfeat{coroutine} was released as an experimental feature in Kotlin 1.1 and it was made stable in Kotlin 1.3 in October of 2019.
Comparing with \rfeat{type alias}, also released in Kotlin 1.1, \rfeat{coroutines} are found in 23\% of applications, whereas \rfeat{type aliases} are found in 13\%.
Since in Java, the concept of \rfeat{type alias} does not exist, \rfeat{type alias} is not interoperable with Java. 
Therefore, this could be one possible reason for this level of adoption.
On the other hand, \rfeat{coroutine} might be used to perform different actions concurrently without block Android's main thread.
For instance, one could replace \textit{AsyncTaks} usage by \rfeat{coroutine} since \textit{AsyncTask} was marked as deprecated~\citep{AsyncTask2019}.

Delegation has been proven to be an alternative to inheritance \citep{fowler2018refactoring}, but we found that \textit{super delegation} was used only in 39 out of \datasetsize~(10\%) applications. 
Moreover, we observed that Kotlin applications have a median of 0 occurrences of \rfeat{super delegation}.
Therefore, we believe that the fact of inheritance is essentially absent in mobile applications~\citep{Minelli2013} can explain this finding. 

Among the least used features, we found two features released in Kotlin 1.3.
Although \rfeat{contract} was released as an experimental feature, we found one application using it and \textit{inline class} was found in 4 applications.
The same number of applications that use \textit{tail-recursive} functions, that was released in Kotlin 1.0.

\begin{tcolorbox}[colback=white, boxrule=0.5pt, boxsep=0.0pt]
\textbf{Finding 4:} 
\textit{Less than 35\% of Android apps have instances of features like, inline function, inline class and tail-recursive function, that might improve their performance~\citep{KotlinInlineFunctions2016,KotlinTail2016,KotlinInlineClasses2018}.
}
\end{tcolorbox}

\begin{tcolorbox}[boxsep=0.0pt]
Response to RQ 1:  \emph{\rqone}

We studied 26 Kotlin features, and as a result, we found that Android developers use all features of them. We identified three groups of features:
\begin{inparaenum}[\it i)]
\item 7 features used in at least 80\% of applications;
\item 9 features used in more than 48\% and less than 80\% of applications;
\item 10 used in less than 33\%.
\end{inparaenum}

Furthermore, we found that \rfeat{type inference}, \rfeat{lambdas} and \rfeat{safe calls}  are the most used features, being found on 98\%, 95\% and 89\% of applications, respectively.

\end{tcolorbox}

\begin{figure}
    \centering
    \includegraphics[width=\columnwidth]{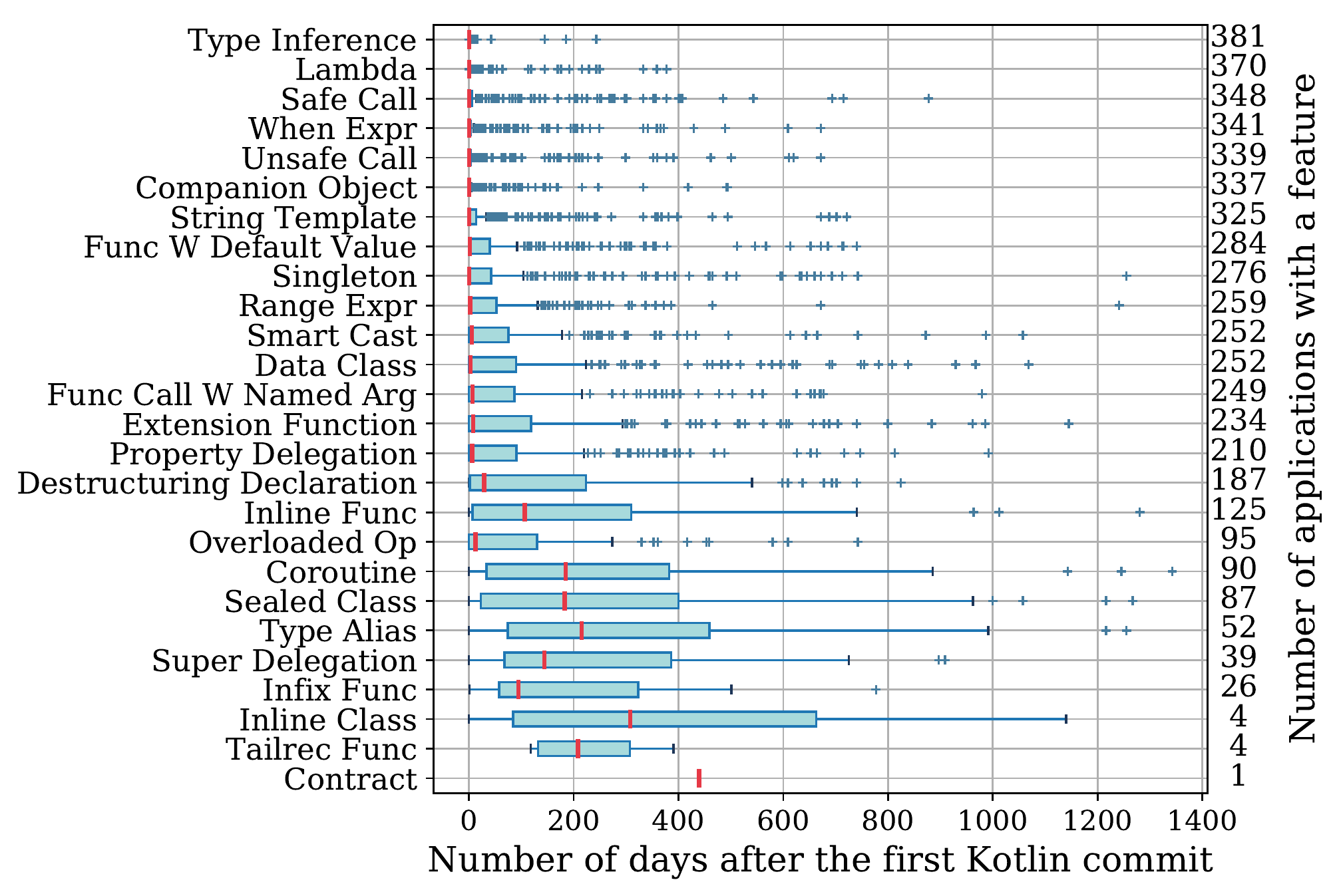}
    \caption{Distributions of the number of days 
    between the first Kotlin commit and the commit that introduces the first instance of a feature.}
    \label{fig:commit_introduce}
\end{figure}


\subsection{RQ2: \rqtwo}
\label{sec:result:rq2}

To answer this research question, we used a metric defined in Section~\ref{sec:method:timeintro}, named \textit{introduction moment}.  
Figure~\ref{fig:commit_introduce} displays its distribution. 
We found that:

\begin{tcolorbox}[colback=white, boxrule=0.5pt, boxsep=0.0pt]
\textbf{Finding 5:} 
\textit{In the following ten days after the first commit with Kotlin code, 15 out of 26 features are added into Android applications.
}
\end{tcolorbox}

However, comparing the introduction moment of the most and the least used features, it turned out distinct behaviors.
We found that the most used features, \textit{type inference, lambda, safe call, when expressions, companion object, unsafe call, string template and singleton}, presented in median the \textit{introduction moment} smaller than 1.
While the least used features presented a high introduction moment.

\begin{tcolorbox}[colback=white, boxrule=0.5pt, boxsep=0.0pt]
\textbf{Finding 6:} 
\textit{The least used Kotlin features tend to be introduced later on the applications' history when compared with the most used features.}
\end{tcolorbox}

Furthermore, the two features with the highest introduction moment, respectively, 308 and 439, were \textit{inline class} and \textit{contract}, both released in Kotlin 1.3.
These features were made available more recently compared to the others. 
This can explain the highest introduction moment.
Among features released in Kotlin 1.1, we noted that \textit{coroutine},  released initially as an experimental feature, presented a smaller introduction moment than \textit{type alias}.
We analyzed the date of the commits that introduced \textit{coroutine} and we found that for 87 out of 90 applications (96\%) added \textit{coroutines} when it was still an experimental feature.
\textit{Contract}, found in one application, it is another experimental feature used by developers.
Therefore, as found by Dyer et al. \cite{Dyer2013} in their study about the usage of Java features, we observed that developers start adopting features early when they are experimental.


\begin{tcolorbox}[boxsep=0.0pt]
Response to RQ 2:  \emph{\rqtwo}
Most features, 15 out of 26, are added into the first days of development using Kotlin.
Moreover, while the most used Kotlin features \textit{type inference, lambda, safe call, when expressions, companion object and string template} tend to be introduced in the first commit with Kotlin code,  the least used features tend to be introduced later into an application. 
\end{tcolorbox}

\subsection{RQ3: \rqthree}

Table~\ref{tab:feature:fit} presents the results of this research question obtained using the methodology presented in Section~\ref{sec:met:fit}.
Each cell shows the number and the percentage of applications whose the evolution of a feature $f$ (a row) is \emph{better described} by trend $t$ (a column). 
Additionally, the number of applications analyzed that have a feature $f$ is displayed in the column Total applications.

For 11 out of 26 (42\%) features studied are better described by \etoneabb, a constant rise trend.
Moreover, 7 out of 26 (6\%) features are better described by \etnineabb.
Other 6 features (23\%) presented the behavior of stability intervals separated by a gradual rise (trend \etsevenabb~Figure~\ref{fig:sample:T9}).
We also observed that 2 (7\%) features better described by \ettenabb.
Note that the feature \textit{tail-recursive function} is better described by two trends, \etnineabb~and~\ettenabb.
Additionally, the \etsevenabb~trend better describes 1 feature (3\%), \textit{inline class}.
Finally, we did not find any feature better described by stability. 

\begin{tcolorbox}[colback=white, boxrule=0.5pt, boxsep=0.0pt]
\textbf{Finding 7:} 
\textit{In general, the number of instances of features tends to grow along with the applications' evolution.}
\end{tcolorbox}

In Table~\ref{tab:feature:fit}, the column \textbf{Inc} shows the total number of applications and the percentage of applications which are better described by any trend whose the number of instances increases along the application's evolution (i.e., \etoneabb, \etfourabb, \etsixabb, \etsevenabb, \etnineabb). 
Analogously, the column \textbf{Dec}, shows the sum of \ettwoabb, \etfiveabb, \eteightabb, \ettenabb.
Furthermore, column \textbf{\etelevenabb}~represents the Instability trend.
Consequently, 100\% of applications are represented by columns \textbf{\etelevenabb}, \textbf{Inc} and \textbf{Dec}.

\bgroup
\def\arraystretch{1.3}
\setlength{\tabcolsep}{0.15cm}

\begin{table*}
\scriptsize
\centering
\caption{This table shows for each Kotlin feature $X$ (rows) and for each evolution formula $f$ (columns) the number of applications where the formula  $f$ better \emph{describes} the evolution of the feature $X$. 
The last row, Total, shows how many features were better described by a trend (column)}.
\begin{tabular}{|p{2.4cm}|| c c c c c c c c c c c||c c||c|}
\hline 
\diagbox[height=0.33cm,width=2.7cm]{Features}{Trends}& \etoneabb & \ettwoabb & \etthreeabb & \etfourabb & \etfiveabb & \etsixabb & \etsevenabb & \eteightabb & \etnineabb & \ettenabb & \etelevenabb & Inc & Dec & Total\\


\hline

Type Inference & \textbf{128 (36\%)} & 3 (1\%) & 0 (0\%) & 60 (17\%) & 1 (0\%) & 53 (15\%) & 10 (3\%) & 0 (0\%) & 5 (1\%) & 4 (1\%) & 91 (26\%) & 256 (72\%) & 8 (2\%) & 355 \\
Lambda & \textbf{124 (36\%)} & 2 (1\%) & 0 (0\%) & 67 (20\%) & 4 (1\%) & 31 (9\%) & 36 (10\%) & 0 (0\%) & 14 (4\%) & 4 (1\%) & 61 (18\%) & 272 (79\%) & 10 (3\%) & 343 \\
Safe Call & \textbf{94 (30\%)} & 2 (1\%) & 0 (0\%) & 64 (20\%) & 3 (1\%) & 38 (12\%) & 33 (10\%) & 1 (0\%) & 21 (7\%) & 9 (3\%) & 51 (16\%) & 250 (79\%) & 15 (5\%) & 316 \\
When Expr & \textbf{85 (29\%)} & 3 (1\%) & 0 (0\%) & 47 (16\%) & 2 (1\%) & 28 (10\%) & 72 (24\%) & 0 (0\%) & 29 (10\%) & 5 (2\%) & 23 (8\%) & 261 (89\%) & 10 (3\%) & 294 \\
Unsafe Call & \textbf{60 (19\%)} & 32 (10\%) & 0 (0\%) & 54 (17\%) & 19 (6\%) & 25 (8\%) & 41 (13\%) & 3 (1\%) & 17 (5\%) & 11 (4\%) & 51 (16\%) & 197 (63\%) & 65 (21\%) & 313 \\
Companion Object & \textbf{75 (26\%)} & 11 (4\%) & 0 (0\%) & 38 (13\%) & 8 (3\%) & 33 (11\%) & 57 (20\%) & 2 (1\%) & 39 (13\%) & 5 (2\%) & 22 (8\%) & 242 (83\%) & 26 (9\%) & 290 \\
String Template & \textbf{80 (27\%)} & 5 (2\%) & 0 (0\%) & 39 (13\%) & 6 (2\%) & 32 (11\%) & 59 (20\%) & 1 (0\%) & 28 (9\%) & 7 (2\%) & 38 (13\%) & 238 (81\%) & 19 (6\%) & 295 \\
Func With Default Value & 50 (21\%) & 3 (1\%) & 0 (0\%) & 41 (18\%) & 4 (2\%) & 19 (8\%) & \textbf{64 (27\%)} & 3 (1\%) & 27 (12\%) & 6 (3\%) & 17 (7\%) & 201 (86\%) & 16 (7\%) & 234 \\
Singleton & 33 (15\%) & 6 (3\%) & 0 (0\%) & 36 (16\%) & 7 (3\%) & 25 (11\%) & \textbf{51 (22\%)} & 3 (1\%) & 29 (13\%) & 24 (11\%) & 13 (6\%) & 174 (77\%) & 40 (18\%) & 227 \\
Range Expr & 30 (14\%) & 14 (7\%) & 0 (0\%) & 28 (13\%) & 5 (2\%) & 17 (8\%) & \textbf{47 (22\%)} & 6 (3\%) & 38 (18\%) & 14 (7\%) & 12 (6\%) & 160 (76\%) & 39 (18\%) & 211 \\
Smart Cast & 38 (18\%) & 14 (7\%) & 0 (0\%) & 35 (17\%) & 8 (4\%) & 20 (10\%) & \textbf{40 (19\%)} & 4 (2\%) & 25 (12\%) & 14 (7\%) & 10 (5\%) & 158 (76\%) & 40 (19\%) & 208 \\
Data Class & 44 (22\%) & 2 (1\%) & 0 (0\%) & 28 (14\%) & 5 (2\%) & 19 (9\%) & \textbf{51 (25\%)} & 0 (0\%) & 35 (17\%) & 6 (3\%) & 14 (7\%) & 177 (87\%) & 13 (6\%) & 204 \\
Func Call With Named Arg & \textbf{48 (22\%)} & 2 (1\%) & 0 (0\%) & 40 (19\%) & 6 (3\%) & 20 (9\%) & 41 (19\%) & 1 (0\%) & 30 (14\%) & 8 (4\%) & 19 (9\%) & 179 (83\%) & 17 (8\%) & 215 \\
Extension Function & \textbf{47 (23\%)} & 4 (2\%) & 0 (0\%) & 32 (16\%) & 7 (3\%) & 23 (11\%) & 42 (21\%) & 1 (0\%) & 22 (11\%) & 7 (3\%) & 16 (8\%) & 166 (83\%) & 19 (9\%) & 201 \\
Property Delegation & \textbf{40 (22\%)} & 11 (6\%) & 0 (0\%) & 32 (18\%) & 8 (4\%) & 22 (12\%) & 23 (13\%) & 3 (2\%) & 18 (10\%) & 5 (3\%) & 17 (9\%) & 135 (75\%) & 27 (15\%) & 179 \\
Destructuring Declaration & 23 (16\%) & 9 (6\%) & 0 (0\%) & 14 (10\%) & 5 (3\%) & 9 (6\%) & \textbf{33 (23\%)} & 4 (3\%) & 25 (17\%) & 16 (11\%) & 5 (3\%) & 104 (73\%) & 34 (24\%) & 143 \\
Inline Func & 11 (12\%) & 5 (5\%) & 0 (0\%) & 11 (12\%) & 5 (5\%) & 5 (5\%) & 17 (18\%) & 2 (2\%) & \textbf{20 (21\%)} & 14 (15\%) & 5 (5\%) & 64 (67\%) & 26 (27\%) & 95 \\
Overloaded Op & 9 (14\%) & 3 (5\%) & 0 (0\%) & 5 (8\%) & 2 (3\%) & 5 (8\%) & 11 (17\%) & 2 (3\%) & \textbf{15 (23\%)} & 11 (17\%) & 1 (2\%) & 45 (70\%) & 18 (28\%) & 64 \\
Coroutine & \textbf{17 (22\%)} & 1 (1\%) & 0 (0\%) & 12 (16\%) & 1 (1\%) & 8 (11\%) & 16 (21\%) & 0 (0\%) & 6 (8\%) & 1 (1\%) & 14 (18\%) & 59 (78\%) & 3 (4\%) & 76 \\
Sealed Class & 6 (9\%) & 3 (5\%) & 0 (0\%) & 10 (15\%) & 1 (2\%) & 5 (8\%) & 7 (11\%) & 0 (0\%) & \textbf{26 (40\%)} & 3 (5\%) & 4 (6\%) & 54 (83\%) & 7 (11\%) & 65 \\
Type Alias & 2 (7\%) & 1 (3\%) & 0 (0\%) & 6 (20\%) & 1 (3\%) & 0 (0\%) & 8 (27\%) & 0 (0\%) & \textbf{10 (33\%)} & 2 (7\%) & 0 (0\%) & 26 (87\%) & 4 (13\%) & 30 \\
Super Delegation & 3 (10\%) & 0 (0\%) & 0 (0\%) & 1 (3\%) & 3 (10\%) & 4 (14\%) & 3 (10\%) & 0 (0\%) & \textbf{10 (34\%)} & 4 (14\%) & 1 (3\%) & 21 (72\%) & 7 (24\%) & 29 \\
Infix Func & 3 (19\%) & 0 (0\%) & 0 (0\%) & 0 (0\%) & 1 (6\%) & 0 (0\%) & 3 (19\%) & 0 (0\%) & 3 (19\%) & \textbf{6 (38\%)} & 0 (0\%) & 9 (56\%) & 7 (44\%) & 16 \\
Inline Klass & 0 (0\%) & 0 (0\%) & 0 (0\%) & 0 (0\%) & 0 (0\%) & \textbf{1 (100\%)} & 0 (0\%) & 0 (0\%) & 0 (0\%) & 0 (0\%) & 0 (0\%) & 1 (100\%) & 0 (0\%) & 1 \\
Tailrec Func & 0 (0\%) & 0 (0\%) & 0 (0\%) & 0 (0\%) & 0 (0\%) & 0 (0\%) & 0 (0\%) & 0 (0\%) & \textbf{1 (50\%)} & \textbf{1 (50\%)} & 0 (0\%) & 1 (50\%) & 1 (50\%) & 2 \\
Contract & 0 (0\%) & 0 (0\%) & 0 (0\%) & 0 (0\%) & 0 (0\%) & 0 (0\%) & 0 (0\%) & 0 (0\%) & \textbf{1 (100\%)} & 0 (0\%) & 0 (0\%) & 1 (100\%) & 0 (0\%) & 1 \\
\hline
Total & 11 & 0 & 0 & 0 & 0 & 1 & 6 & 0 & 7 & 2 & 0 & - & - & - \\
\hline
\end{tabular}
\label{tab:feature:fit}
\end{table*}

\egroup

Now we explain our results for the three most used Kotlin features 
, \rfeat{type inference}, \rfeat{lambda}, and \rfeat{safe call}. 
We found that the majority of applications presented a behavior of increasing the number of instances along the applications' evolution. 
Table~\ref{tab:feature:fit} shows this behavior in 72\% of applications containing \rfeat{type inference}, 79\% of applications containing \rfeat{lambda} and 79\% of applications containing \rfeat{safe calls}.
Moreover, the constant rise trend, \etoneabb, better described the evolution of these features in 36\%, 36\% and 30\% of applications, respectively.
Considering the usage of \rfeat{type inference}, we found 91 (26\%) applications whose number of instances varies between increase and decrease intervals along with applications' evolution. 
Furthermore, we observed only 8 (2\%) applications whose number of \rfeat{type inference} instances decreases during applications' evolution. 

\begin{tcolorbox}[boxsep=0.0pt]
Response to RQ 3:  \emph{\rqthree}
Developers tend to add more instances along the evolution of Android applications of 24 out of 26 (92\%) features studied. 

\end{tcolorbox}

Finally, as we described in Section \ref{sec:met:trend}, to identify the trend that better describes the evolution trend of a feature, we used R-squared, which measures how close the data are to the fitted formula.
The R-squared assumes values between 0 and 1, and 1 means a perfect fitting. 
In our experiment, the median R-squared was 0.88, which means that 50\% of the selected formulas describe almost perfectly the evolution trends.
Moreover, we found that 75\% of evolution trends fitted presented an R-squared greater than 0.74, and only outliers have R-squared values lower than 0.43.

\subsection{Implications}

As a consequence of our study, we defined several implications that we group according to the needs of researchers, tool builders, and software developers. 

\subsubsection{Researchers}\mbox{} \\
The findings from RQ1 reveal some novel research opportunities:

\textit{Lambda} is the second most used feature and with a median of 71 instances per application. This finding reveals that Android developers rely on \textit{lambda} to build their applications. \citet{Mazinanian:2017:UUL:3152284.3133909} studied the usage of \textit{lambda} in Java and found several misuse cases. Therefore, one could perform a similar study considering the use of \textit{lambda} in Kotlin.
Another possibility is to investigate how different is the usage of \textit{lambda} in mobile applications
compared with non-mobile applications written in Java and Kotlin.

\textit{Safe call} and \textit{Unsafe call} are used in the majority applications written in Kotlin. 
While \textit{safe call} can protect applications of \textit{NullPointerException (NPE)}, \textit{unsafe call} makes them more vulnerable. 
The study performed by \citet{Coelho2015} have shown that considering uncaught exceptions caused by errors in programming logic in Android applications, the NPE is the most commom exception. Therefore, researchers can further explore if the usage of \textit{safe call} prevents  the occurrence of these exceptions effectively and, on the contrary, if the usage \textit{unsafe call} impact negatively.

\subsubsection{Tool builders and IDE designers}\mbox{} \\
Android developers rely on lambdas expressions, which might result in a more concise code when compared with the approach that uses inner classes. 
However, a large number of instances can result in overuse.
Therefore, IDE can warn developers about overused lambda, for instance, when a high number of lambdas are nested.

We found a median of 16 occurrences of \textit{unsafe calls} per application that includes some situations where developers could use Kotlin's built-in functions instead. 
Considering that, removing unnecessary instances of \textit{unsafe calls} could improve the null-safety of Android applications, automated code assistance could suggest this type of refactoring.

RQ1 highlights that less than 35\% of Android applications have instances of \textit{inline function}, \textit{inline class} and \textit{tail recursive}, although accurate use of them could improve applications' performance. 
Therefore, IDEs could detect refactoring scenarios and recommend more efficient implementations.

\subsubsection{Developers}\mbox{} \\
Regarding the use of \textit{named arguments}, we consider that developers could write a more idiomatic code following the Kotlin conventions that suggest the use of named argument when a method takes multiples parameters of the same primitive type or for parameters of Boolean type~\citep{kotlin_conventions}.
IDEs could identify spots where \textit{named argument} could be applied in order to follow the Kotlin conventions. On the contrary, we noted that Android Studio renders the name of arguments in functions calls when they are not provided, resulting in more readable code. However, when someone tries to read that code outside the IDEs, for instance, to perform a code review on GitHub, the readability is not the same as in the IDE.

We observed that less than 1\% of properties defined in Kotlin applications are delegated. Inspecting the applications' code, we found opportunities to use property delegation to make a more concise code. 
As suggested in the official Android development documentation, property delegation could save development time in some scenarios~\citep{AndroidCommonPatterns}. 
Better tutorials and training material could help to promote this feature. 

In RQ2, we found that most features are added into applications in the following ten days after the first commit with Kotlin code.
As pointed by~\citet{Tufano2015}, 
most of the times code artifacts are affected by bad
smells since their creation. 
Therefore, a solid understanding of these features would help developers to use them correctly since the initial phase of development, potentially improving the code quality.
A dedicated training focused on these features would especially help novice Kotlin developers.
On the contrary, the least used Kotlin features tend to be introduced later than the most used features.
A possible reason is that developers need more experience with Kotlin to start using these features.
Then, promoting these features could help developers to identify opportunities for using them in the early days of development.

We observed that developers add more instances along with the applications’ evolution.
As the number of instances grows, applications become more prone to overusing of features.
Continuous integration and static analyzers could be used to track the use of features and to identify misuse and anti-patterns.





\section{Threats to Validity}
\label{sec:threatsvalidity}

\subsection{Internal}

\textit{Selection of features.} 
We investigated the adoption and usage evolution of a subset of Kotlin features not available in Java.
As the criterion used to select the targeted features impact our results, we consider the ones listed in the official Kotlin documentation to avoid any bias~\citep{KotlinComparison2016}.

\textit{Feature identification.} Our results depend on how precisely we identify the target features.
Due to the absence of a benchmark of the use of Kotlin features, To evaluate our tool, we manually analyzed 96 instances of each feature and we found 100\% of precision.

\textit{Evolution trend identification.} One of the goals of our work is to identify evolution trends.
To avoid any bias,
we established 11 evolution trends (See Section~\ref{sec:met:trends}) similarly used in other studies~\citep{GoisMateus2019,Hecht2015,Malavolta2018}.
Moreover, to match a feature evolution with one of the studied trends, we defined 6 formulas.
However, it is possible that other formulas, not used in this paper, fit better with any evolution trend.

\subsection{External}

\textit{Representativeness of FAMAZOA. }Our work relies on FAMAZOA~\citep{GoisMateus2019}, a dataset of open-source mobile applications written in Kotlin.
Considering the number of applications published on Google Play, FAMAZOA represents a small parcel since it only contains open-source applications, limiting the generalization of our findings.
However, to the best of our knowledge, it is the largest dataset of Android open-source applications written in Kotlin.

\textit{Developer's experience. }In our work, we analyzed the source code of different Android applications.
The use of some studied features may require more experience from developers to be used properly. 
However, we did not consider information about developers' experience.
Nevertheless, all applications analyzed were published on F-droid or Google Play, then it represents a current snapshot of Android development. 
We will consider this aspect in a future work.

\section{Related Work}
\label{sec:relatedwork}
We group the related work into three areas:  \begin{inparaenum}[1)] 
\item studies on adoption of language features
\item studies on the usage evolution of language features and
\item evolution of Android applications.
\end{inparaenum}

\textit{Studies on adoption of language features. }
Several empirical studies about features of different programming languages have been done. 
\citet{Sutton2010} developed a tool to identify the use of generic libraries in C++ projects to support a programmer’s comprehension. 
\citet{Parnin2011,Parnin2013} conducted an empirical study to understand how Java generics have been integrated into open-source software by mining repositories of 40 popular Java programs.
They found that one or two contributors often adopt generics and generics can reduce the number of typecasts in a project.
\citet{Uesbeck:2016:ESI:2884781.2884849} investigated the impact of lambda expressions on the development, debugging, and testing effort in C++.
\citet{Pinto2016} studied the energy efficiency of Java's Thread-Safe Collections and concluded that simple design decisions could impact energy consumption.
\citet{Chapman:2016:ERE:2931037.2931073} studied the usage of regular expressions in Python conducting an empirical study of open-source projects from GitHub.
They found six common behaviors that describe how regular expressions are often used in practice.
JavaScript features were also the target of studies. 
~\citet{silva2015does} developed a tool to investigate how class emulation is employed in JavaScript applications. They observed that 26\% of 50 most popular applications from Github do not use class.
\citet{Gallaba2017} focused on the usage of \textit{asynchronous callbacks} in JavaScript. They evaluated their tool, PROMISESLAND, and found that it substitutes callbacks to promises (an alternative to asynchronous callbacks), correctly. Moreover, none of them has studied the features of Kotlin.

\textit{Studies on the usage evolution of language features. }
Various studies on the adoption of features over time have been done. \citet{Pinto:2015:LSU:2794082.2794114} and \citet{Wu:2016:EES:2949070.2949129} performed large empirical studies to investigate the usage of concurrency constructs in open-source applications in Java and C++, respectively.
\citet{Pinto:2015:LSU:2794082.2794114} found more than 75\% of 2227 projects either explicitly create threads or employ some concurrency control mechanism.
\citet{Wu:2016:EES:2949070.2949129} found that small-size applications introduce concurrency constructs more intensively and quicker than medium-size applications and large-size applications. 
\citet{Osman2017,Osman2017MSR} conducted empirical studies on long-lived Java systems and observed that the amount of error-handling code, the number of custom exceptions and their usage in catch handlers and throw statements increase as projects evolve.
\citet{yu:hal-02091516} conducted an empirical study to evaluate the usage, evolution and impact of Java Annotations.
Malloy and Power \citep{Malloy2017,Malloy2019} investigated the degree to which Python developers are migrating from Python 2 to 3 by measuring the adoption of Python 3 features.
They found that developers were choosing to maintain backward compatibility with Python 2 instead of exploiting the new features and advantages of Python 3. 
To identify the uses over time of new features of Java, \citet{Dyer:2014:MBA:2568225.2568295} analyzed \numprint{23000} open-source Java projects. 
They observed that all features are used even before their release date and that there were still millions of opportunities for use. 
\citet{Mazinanian:2017:UUL:3152284.3133909} focused on the evolution trend of one Java feature, Lambdas.
They conducted an empirical study, statically analyzing the source code of 241 open-source projects and interviewing 97 developers who introduced lambdas in their projects. 
As a result, they revealed an increasing trend in the adoption of lambdas.
In contrast with the previous studies, our work focuses on a set of Kotlin features, their adoption and evolution along with applications' version.

\textit{Evolution of Android applications. }
Several aspects of the evolution of Android applications were investigated in the literature. 
\citet{Hecht2015} presented an approach for performing
static code analysis on Android applications' bytecode to detect code smells and they identified five different quality evolution trends.
\citet{Calciati:2017:AEP:3104188.3104195} investigated the evolution of permission requests analyzing more than \numprint{14000} applications. 
They found that applications tend increase the number permissions needed over time.
\citet{GoisMateus2019} conducted the first empirical study about Android applications written in Kotlin.
Analyzing the source code of Android applications they identified 12 evolution trends.
\citet{Malavolta2018} investigated the evolution of 6 maintainability issues along with the evolution of Android applications.
Inspecting the density of maintainability issues, they identified 12 different evolution trends and concluded that independently from the type of development activity and notwithstanding the issue type, maintainability issue density grows until it stabilizes.
\citet{habchi:hal-02059097} presented the first large-scale empirical study that investigates the survival of eight types of Android code smells. 
They reported that while in terms of time Android code smells can remain in the codebase for years before being removed, it only takes 34 effective commits to remove 75\% of them. 
\citet{Calciati:2018:DRC:3196398.3196449} proposed a framework to analyze the evolution of Android applications.
Analyzing 235 applications with at least 50 releases, they found that Android applications tend to have more leaks of sensitive data over time.
These studies tie into our research because they focus on the evolution of Android applications.
Our study applied a methodology similar to~\citep{Hecht2015,GoisMateus2019,Malavolta2018} to identify evolution trends of feature usage.
However, these studies used a manual approach to classify the identified trends.
Differently, in our work, we applied an automated classification approach. 
To the best of our knowledge, this work is the first empirical study about the adoption and evolution of Kotlin features.

\section{Conclusion}
\label{sec:conclusion}


We conducted the first empirical study exploring different aspects of the adoption and evolution of Kotlin features not available in Java for Android development.
We extracted 26 features from applications' source code and we identified when they were used for the first time.
To classify the evolution trends of these features, we first defined 11 evolution trends modeled by 5 functions. Then, we created an automated method of classification.
We found that 15 out of 26 features are used by the majority of Android applications.
\rfeat{Type inference}, \rfeat{lambda} and \rfeat{safe call} are the most used features, they are presented in 98\%, 95\% and 89\% of applications, respectively.
Furthermore, we observed that the most used features are added in the first Kotlin commit, and  the least used features are introduced in the latest commits. Finally, we found that 24 out of 26 features presented an increasing of features instances along with the applications' evolution.
As future work, we intent to perform a qualitative study that interviews developers focused on understanding the motivation behind the use of Kotlin features to complement this work. 


\bibliographystyle{ACM-Reference-Format}
\bibliography{references}

\end{document}